\documentclass [11pt]{article}
\usepackage{amsmath,amsthm,amsfonts,amscd,eucal,latexsym,amssymb}
\def\sp{\hskip -5pt}
\def\spa{\hskip -3pt}

\newsymbol\bt 1202           
\def\bC{{\mathbb C}}           

\def\bN{{\mathbb N}}

\def\bR{{\mathbb R}}

\newsymbol\rest 1316         

\def\phi{\varphi}

\begin{document}


\hfill{\sl preprint - UTM 601}
\par
\bigskip
\par
\rm


\par
\bigskip
\LARGE
\noindent
{\bf Comments on the Stress-Energy Tensor Operator in Curved Spacetime.}
\bigskip
\par
\rm
\normalsize



\large
\noindent {\bf Valter Moretti}

\large
\smallskip

\noindent
Department of Mathematics,
Trento University,
I-38050 Povo (TN), Italy.\\
E-mail: moretti@science.unitn.it\\

\large
\smallskip

\rm\normalsize



\par
\bigskip
\par
\hfill{\sl September 2001}
\par
\medskip
\par\rm

    \noindent   {\em To my wife Francesca and my daughter Bianca.}   \\

\noindent
{\bf Abstract:}
The technique based on a $*$-algebra of Wick products of field operators
in curved spacetime, in the local covariant version proposed by Hollands
and Wald, is strightforwardly generalized in order to define the stress-energy tensor 
operator in curved globally hyperbolic spacetimes. In particular, the locality and covariance 
requirement is generalized to Wick products of  differentiated quantum fields.
Within the proposed formalism, there is room to accomplish
all of physical requirements provided that known problems concerning the conservation of
the stress-energy tensor 
 are assumed to be related to the interface between quantum and classical
formalism. The proposed stress-energy tensor operator turns out to be conserved and 
reduces to the classical form if field operators are replaced by classical fields
satisfying the equation of motion.
The definition is  based on the existence of convenient counterterms given by
certain local Wick products of differentiated fields. These terms are independent from the arbitrary
length scale (and any quantum state) and they classically vanish on solutions of
Klein-Gordon equation. Considering the averaged stress-energy tensor with respect to Hadamard 
quantum states,
the presented definition turns out to be equivalent to an improved point-splitting renormalization
procedure which makes use of the nonambiguous part of the Hadamard parametrix only that
is determined by the local geometry and the parameters which appear in the Klein-Gordon operator.
In particular, no extra added-by-hand term $g_{\alpha\beta}Q$
and no arbitrary smooth part of the Hadamard parametrix (generated by some  arbitrary smooth  term 
"$w_0$") are involved. The averaged stress-energy tensor obtained  by the  point-splitting procedure  
also coincides with that found by employing the  local $\zeta$-function approach whenever that 
technique can be implemented.

\section{Introduction.}

 In \cite{bfk,bf,hw} the issue is addressed concerning the  definition of  Wick products of
 field operators (and time-ordered products of field operators)  in curved spacetime
 and remarkable results are found (see Section {\bf 3}). The general goal is the definition of the 
perturbative $S$-matrix
  formalism  and corresponding renormalization techniques for self-interacting quantum fields in 
curved spacetime.
  The definition proposed by Hollands and Wald in \cite{hw} also assumes some {\em locality and 
covariance} requirements
  which (together with other properties) almost completely determine {\em local} Wick products. 
Some of the results on Wick polynomials algebra  presented in \cite{hw}
  are straightforward generalizations of Minkowski-spacetime results obtained by D\"utsch and 
Fredenhagen in \cite{df}.
  A more general  approach based on locality and covariance is presented in \cite{bfv}.
  Using the machinery introduced in \cite{hw}, a stress-energy tensor {\em operator} could be 
defined, 
  not only its formal averaged value (also see \cite{bfk} and comments in \cite{bf}, where another  
definition of
  stress-energy operator was  proposed in terms of a different definition of Wick products).  
However, the authors
  of \cite{hw}
  remark that such a stress-energy operator would not satisfy the conservation requirement.

  This paper is devoted to show that, actually,  a natural  (in the sense that it coincides
  with the classical definitions whenever operators are replaced by classical fields) definition
  of a well-behaved stress-energy tensor operator may be given using  nothing but
  local Wick products defined by Hollands and Wald, provided one considers the
  pointed-out problem as due to the interface between classical and quantum formalism.

 The way we follow is related to the attempt to overcome some known drawbacks which arise when 
one tries
 to  define
 a natural {\em point-splitting} renormalization procedure for the stress-energy tensor averaged  with 
respect to
 some quantum state. Let us illustrate these well-known drawbacks \cite{bd,wald94}.

Consider a  scalar real field $\phi$ propagating in a globally hyperbolic spacetime $(M,{\bf g})$.
Assume that the field equations for $\phi$ are linear and induced by some Klein-Gordon
operator $P=-\Delta +\xi R +m^2$ and let  $\omega$ be a quntum state of the quantized field $\hat 
\phi$.
A widely  studied  issue  is the definition of  techniques which compute  {\em averaged} (with respect 
to $\omega$)
products of pairs of field operators $\hat\phi$ evaluated at the same event $z$.
In practice, one is  interested in formal objects like
$\langle{\hat\phi}(z){\hat\phi}(z)\rangle_\omega$.
The {\em point-splitting procedure} consists of replacing classical terms $\phi(z)\phi(z)$ 
by some argument-coincidence limit of a integral kernel representing a suitable quantum two-point 
function of $\omega$.
A natural choice involves the Hadamard two-point function\footnote{The use of the Hadamard 
function rather than the (Wightman) two-point functions is a matter of  taste, since the final result does not depend 
 on such 
a  choice as a consequence of the bosonic commutation relations.},
$G^{(1)}_\omega(x,y)$, which
 is regular away from light-related arguments for {\em Hadamard states} (see {\bf 2.2}).
The cure for ultraviolet divergences which arise performing the argument-coincidence limit
consists of subtracting the "singular part" of   $G^{(1)}_\omega(x,y)$, (see {\bf 2.2}),
 before taking the coincidence limit $(x,y) \to (z,z)$.
  This is quite a well-posed procedure if the state is Hadamard since, in that case, the
 singular part of the two-point functions is known by definition and  is almost completely determined
 by the geometry and the K-G operator. 
 The use of such an approach for
 objects involving derivatives of the fields, as the stress-energy tensor, turns out to be more 
problematic.
 The na\"\i ve point-splitting
 procedure consists of the following limit
  $$\langle \hat{T}_{\mu\nu}(z) \rangle_\omega = \lim_{(x,y) \to (z,z)} {D}_{\mu \nu}(x,y) 
[G_\omega^{(1)}(x,y) -
  (Z_n(x,y) + W(x,y))]$$
 where  $Z_n$ is the expansion of the singular part of $G_\omega^{(1)}$ in powers of 
 the squared geodesic distance $s(x,y)$ of $x$ and $y$ truncated at some sufficiently large order $n$,
 and ${D}_{\mu\nu}(x,y)$ is a non-local differential operator obtained by {\em point-splitting}  the
 the form of the stress-energy tensor \cite{bd,wald94} (see  (\ref{Tclassic}) in {\bf 2.1}).
  $W$ is an added smooth function of $x,y$.\\
That procedure  turns out to be plagued by several drawbacks  whenever $D=dim(M)$ is even 
($D=4$ in particular).
  Essentially, (a) the produced averaged stress-energy tensor turns out not to be conserved 
  (in contrast with Wald's axioms on stress-energy
  tensor renormalization \cite{wald94})
  and (b)  it does not take  the {\em conformal anomaly} into account \cite{wald94} which also arises 
employing
 different renormalization approaches \cite{bd}.
  (c) The choice of the term $W$ turns out to be quite messy.
   Indeed, a formal expansion of  $W$  is known in terms of powers of the squared geodesic distance 
\cite{wald78},
   but it is completely determined {\em only if} the first term $W_0(x,y)$ of the expansion is given. 
   However, it seems that  there is no completely determined natural choice for $W_0$  (see 
discussion
 and references in \cite{wald78,wald94,fulling}). It is not possible to drop the term $W$ if $D$ is 
even. 
 indeed, an arbitrary length scale $\lambda$ is necessary in the 
 definition of $Z_n$ and changes of $\lambda$ give rise to an added term $W$.
  As a minor difficulty we notice that (d) important results concerning  the issue of the conservation
  of the obtained stress-energy tensor \cite{wald78} required both the  analyticity  of the manifold
  and  the metric  in order to get  convergent expansions for the singular part of $G^{(1)}_\omega$. \\
  The traditional cure  for (a) and (b) consists
  of {\em by hand}  improving the prescription  as
  \begin{align}
  \langle \hat{T}_{\mu\nu}(z) \rangle_{\omega} = \lim_{(x,y) \to (z,z)} {D}_{\mu 
\nu}(x,y)[G_\omega^{(1)}(x,y) - (Z_{n}(x,y) + W(x,y))]
  + g_{\mu\nu}(z)Q(z)
         \label{pppro'}  \:,
 \end{align}
  where  $Q$ is a suitable scalar function of $z$ determined by imposing the
  conservation of the final tensor field. {\em A posteriori},
 $Q$ seems to be determined by the geometry
  and $P$ only.

  Coming back to the stress-energy {\em operator},  one expects that  any  conceivable definition
  should produce results in agreement  with the point-splitting renormalization procedure, whenever 
one takes
  the averaged value of that operator with respect to  any Hadamard state $\omega$. However, the 
appearance  of the term
  $Q$ above could not allow a definition in terms of local Wick products of field operators only.

  In Section 2 we prove  that it is possible to "clean up" the point-splitting procedure. In fact, we 
suggest
   an improved procedure which, preserving all of the
 relevant physical results, is not affected by the drawbacks pointed out above. In particular, it does 
not
 need  added-by-hand terms as $Q$, employing only mathematical objects completely determined by 
the
 local geometry and the operator $P$. The ambiguously determined term $W$ (not only the first
 term $W_0$ of its expansion) is dropped, barring
 the part depending on $\lambda$ as stressed above.
 Finally, no analyticity assumptions are made. Our prescription can be said  ``minimal'' in the sense 
that it uses
 the local geometry and $P$ only. The only remaining ambiguity is a length scale $\lambda$.
 We also show that the presented prescription produces the same
 renormalized stress-energy tensor obtained by other definitions based on the Euclidean 
functional integral
 approach.

 In Section 3 we show that the improved procedure straightforwardly
 suggests a natural form of the stress-energy  tensor operator
 written in terms of  local Wick products of operators which generalize
 those found in \cite{bf,hw}. This operator is conserved, reduces to the usual classical form, 
whenever 
 operators are replaced by classical fields satisfying the field equation, 
 and  agrees with the point-splitting result if one takes the averaged value with
 respect to any Hadamard state.  To define the stress-energy tensor operator as an element of a 
suitable
 $*$-algebra of formal operators smeared by functions of  ${\cal D}(M)$, we need to further develop 
the
 formalism introduced in \cite{hw}. This is done in the third section, where we generalize  
 the notion of local Wick products given in \cite{hw} to {\em differentiated} local Wick products
proving 
 some technical propositions.

 Concerning notations and conventions, throughout the paper a spacetime, $(M,{\bf g})$, is  a 
connected
 $D$-dimensional smooth  (Hausdorff, second-countable) manifold with
 $D\geq 2$ and equipped with smooth Lorentzian metric ${\bf g}$ (we adopt the signature $-
,+\cdots,+$).
  $\Delta$ denotes the Laplace-Beltrami-D'Alembert operator on $M$, locally given by 
$\nabla_\mu\nabla^\mu$, $\nabla$
  being the Levi-Civita covariant derivative associated with the metric
 ${\bf g}$.
 A spacetime is supposed to be {\em oriented}, {\em time oriented}
 and in particular {\em globally hyperbolic} (see the Appendix.A and \cite{wald84}), also if those 
requirements 
 are not explicitly stated.
 Throughout $\mu_{\bf g}$ denotes the natural positive measure induced
 by the metric on $M$ and  given by $\sqrt{-g(x)} \: dx^1\land \cdots \land dx^n$ in each coordinate 
patch. 
 The divergence, $\nabla \cdot T$,  of a tensor field $T$ is 
  defined  by $(\nabla \cdot T)^{\alpha\ldots}\:_{\beta\ldots} =
 \nabla_\mu T^\mu\:^{\alpha\ldots}\:_{\beta\ldots} =  \nabla^\mu 
T_\mu\:^{\alpha\ldots}\:_{\beta\ldots}$
 in each coordinate patch. Finally, throughout  the paper, ``smooth'' means $C^\infty$.

 \section{Cleaning up the Point-Splitting Procedure.}

 {\bf 2.1.} {\em Classical framework.}
 Consider a smooth  real scalar classical field  $\phi$  propagating  in a smooth $D$-
dimensional
 globally hyperbolic spacetime $(M, {\bf g})$.
 $P\phi = 0$ is  the equation of motion of the field
 the Klein-Gordon operator $P$ being
 \begin{eqnarray}
  P \stackrel {\mbox{\scriptsize  def}} {=}    
-\Delta + \xi R(x)+ V(x) \stackrel {\mbox{\scriptsize  def}} {=} -\Delta + m^2 + \xi R(x) + V'(x) 
\label{operator}\:,
  \end{eqnarray}
 where $\xi\in \bR$ is a constant, $R$ is the scalar curvature, $m^2\geq 0$ is the mass of the field 
and
 $V': M\to \bR$ is any smooth function.
 The symmetric stress-energy tensor,  obtained by variational derivative with respect to the metric 
of  the action\footnote{For $\xi=1/6$, in (four dimensional) Minkowski spacetime and on  solutions
of the field equations, this tensor
coincides with  the so called ``new improved'' stress-energy
tensor \cite{CCJ}.} \cite{wald84},
 reads
\begin{eqnarray}
T_{\alpha\beta}(x) &=&
{\nabla}_\alpha\phi(x){\nabla}_{\beta}
\phi(x) - \frac{1}{2}g_{\alpha\beta}(x)\left({\nabla}_\gamma \phi(x) {\nabla}^{\gamma}\phi(x)
 + {\phi^2(x)}V(x)  \right) \nonumber\\
&+&\xi\left[\left( R_{\alpha\beta}(x) -\frac{1}{2}g_{\alpha\beta}(x) R(x)\right)
 + g_{\alpha\beta}(x) \Delta
- \nabla_\alpha\nabla_{\beta} \right] {\phi^2(x)}\:.
\label{Tclassic}
\end{eqnarray}
 Concerning the ``conservation relation'' of $T_{\alpha\beta}(x)$,  if $P\phi=0$, a direct
computation leads to
\begin{eqnarray}
\nabla^\alpha T_{\alpha\beta}(x) =
-  \frac{1}{2}\phi^2(x) \nabla_{\beta} V'(x)  \label{conservationT}\:.
\end{eqnarray}
It is clear that the right-hand side vanishes  provided $V'\equiv 0$ and (\ref{conservationT}) reduces 
to the proper
conservation relation.
The trace of the stress-energy tensor can  easily be computed
in terms of ${\phi^2(x)}$ only.
In fact, for $P\phi=0$, one finds
\begin{eqnarray}
g_{\alpha\beta}(x){T}^{\alpha\beta}(x) &=& \left[\frac{\xi_D-\xi}{4\xi_D-1}
 \Delta  - V(x) \right] \phi^2(x)
 \label{trace}
\end {eqnarray}
 where $\xi_{D} = (D-2)/[4(D-1)]$ defines the {\em conformal coupling}: For $\xi=\xi_D$, if 
$V\equiv 0$ and $m=0$,
the action of the field $\phi$ turns out to be
 invariant under local conformal transformations 
 (${\bf g}(x) \to \lambda(x) {\bf g}(x)$, $\phi(x) \to \lambda(x)^{1/2 -D/4} \phi(x)$) and
the trace of $T_{\alpha\beta}(x)$ vanishes on field solutions by (\ref{trace}).\\

\noindent {\bf 2.2.} {\em Hadamard quantum states and Hadamard  parametrix.}
From now on ${\cal A}(M,{\bf g})$ denotes the abstract $*$-algebra with unit $1$ generated by 
$1$ and 
the abstract field operators $\phi(f)$ smeared by the functions of ${\cal D}(M) := 
C_0^\infty(M,\bC)$.
The abstract field operators enjoy the following  properties where $f,h\in {\cal D}(M)$ and
$E$ is the advanced-minus-retarded fundamental solution of $P$ which exists in globally hyperbolic 
spacetimes \cite{kw}.\\
 {\bf (a) Linearity}: $f\mapsto {\phi}(f)$ is linear,\\ {\bf (b) Field equation}: ${\phi}(Pf)=0$,\\
{\bf (c) CCR}:  $[\phi(f), \phi(h)] = E(f\otimes h) 1$, $E$ being the {\em advanced-minus-retarded} bi-solution \cite{kw},\\  {\bf (d) Hermiticity}: ${\phi}(\overline{f}) = 
{\phi}(f)^*$.

An algebraic quantum state $\omega: {\cal A}(M,{\bf g}) \to \bC$ on ${\cal A}(M,{\bf g})$ is a 
linear functional
which is  normalized  ($\omega(1)=1$) and positive ($\omega(a^*a)\geq 0$ for every $a\in  {\cal 
A}(M,{\bf g})$).
The GNS theorem \cite{haag} states that there is a triple  $({\cal H}_\omega,\Pi_\omega, 
\Omega_\omega)$ 
associated with $\omega$. ${\cal H}_\omega$ is a Hilbert space  with scalar product 
$\langle,\rangle_\omega$.
$\Pi_\omega$ is a  $*$-algebra representation 
of ${\cal A}(M,{\bf g})$ which takes values in a $*$-algebra of unbounded operators 
defined on the dense invariant linear subspace ${\cal D}_\omega \subset {\cal 
H}_\omega$\footnote{The involution 
being the adjoint conjugation on ${\cal H}_\omega$ followed by the restriction to ${\cal 
D}_\omega$.}. 
The distinguished vector $\Omega_\omega\in {\cal H}_\omega$ satisfies  both 
$\Pi_\omega({\cal A}(M,{\bf g})) \Omega_\omega = {\cal D}_\omega$
and
$\omega(a) = \langle\Omega_\omega, \Pi_\omega(a) \Omega_\omega\rangle_\omega$ for every $a\in 
{\cal A}(M,{\bf g})$.
 Different GNS triple associated to the same state are unitary equivalent. 
 From now on, $\hat{\phi}_\omega(f)$ denotes the closeable field
operator $\Pi_\omega(\phi)$ and  ${\cal A}_\omega(M,{\bf g})$ denotes the $*$-algebra 
$\Pi_\omega({\cal A}(M,{\bf g}))$. 
Wherever it does not produce misunderstandings we  write $\hat \phi$ instead of
 $\hat{\phi}_\omega$ and $\langle, \rangle$ instead of $\langle,\rangle_\omega$.

The {\em Hadamard} two-point function of $\omega$ is defined by
\begin{eqnarray}
G_{\omega}^{(1)}  \stackrel{\mbox{\scriptsize  def}} {=} Re \:\:G_\omega^{(+)}   
\label{hadamard}  \:,
\end{eqnarray}
 $G_\omega^{(+)}$ being  the {\em two-point} function of $\omega$, i.e., the linear map on ${\cal D}(M)\times {\cal D}(M)$ 
\begin{eqnarray}
G_{\omega}^{(+)} &:&  f\otimes g \to \omega({\phi}(f){\phi}(g)) =
\langle \Omega_\omega, \hat{\phi}(f)\hat{\phi}(g)  \Omega_\omega\rangle   
 \nonumber  \:.
\end{eqnarray}

We also assume
that $\omega$ is {\em globally Hadamard} \cite{kw,wald94,ra}, i.e., it satisfies the\\
{\bf Hadamard requirement}: {\em $G_{\omega}^{(+)} \in {\cal D}'(M\times M)$ and takes the 
 singularity structure of (global) Hadamard form
  in a causal normal neighborhood $N$ of a Cauchy surface $\Sigma$ of $M$}.\\
   In other words, for  $n=0,1,2,\ldots$, the distributions
     $G_{\omega}^{(+)} - \chi Z^{(+)}_{n}\in {\cal D}'(M\times M)$,
can be  represented by functions of  $C^n(N\times N)$. 
 $Z^{(+)}_{n}$
is the {\em Hadamard parametrix} truncated at the order $n$
 and defined on test functions supported in  $C_z\times C_z$ for every $z\in M$, $C_z$
 being a convex normal neighborhood 
of $z$ (see the Appendix A and \cite{kw,sv} 
for the definition of $\chi$ and $N$). 
Since we are interested in the local behavior of the distributions we ignore the smoothing fuction $\chi$
in the following because $\chi(x,y)=1$ if $x$ is sufficiently close to $y$. 
The {\em propagation} of the global Hadamard structure in the whole spacetime \cite{fsw} (see also \cite{kw,wald94,sv})  entails 
the independence of the definition of Hadamard state from $\Sigma$, $N$ and $\chi$. 
It also implies that $G_{\omega}^{(1)}$ (as well as $G_\omega^{(+)}$)
is a smooth function, $(x,y) \mapsto G_{\omega}^{(1)}(x,y)$
away from the subset  of  $M\times M$ made of the pairs of points $x,y$ such that either $x=y$ or they are 
 light-like related.
If $C_z$ is a convex normal neighborhood of $z$, using Hadamard condition and 
the content of the Appendix A, one
proves that $Re(Z^{(+)})$ is represented by a smooth kernel $Z(x,y)$ if $s(x,y)\neq 0$ and 
the map
$(x,y)\mapsto G^{(1)}_\omega(x,y) - Z_{n}(x,y)$ can be continuously extended into a function of $C^n(C_z\times C_z)$.
For $s(x,y) \neq 0$,
\begin{align}
 Z_{n} (x,y) \:\:&= \:\:\beta^{(1)}_D \: \frac{{U}(x,y) }{{s}^{D/2-1}(x,y)}+
 \beta^{(2)}_D {V}^{(n)}(x,y) \: \ln
 \frac{|s(x,y)|}{\lambda^2}&\:\:\:\:\:\:   \mbox{if $D$ is even},   \label{Z2'}    \\
 {Z}_{n} (x,y) \:\:&= \:\: \beta^{(1)}_D \theta(s(x,y))
 \frac{{T}^{(n)}(x,y) }{s^{D/2-1}(x,y)} &\:\:\:\:\:\:   \mbox{if $D$ is odd}  \:.   \label{Z2''}
\end{align}
$\theta(x) =1$ if $x>0$  and $\theta(x)=0$ otherwise.
 The smooth real-valued functions $U,V^{(n)},T^{(n)}$ are defined
by recursive (generally divergent) expansions in powers of  the (signed) squared geodesical distance 
$s(x,y)$ and are completely determined by the metric and the operator $P$. 
$\beta_D^{(i)}$ are numerical coefficients. $\lambda>0$ is an arbitrarily fixed length scale.
    Details are supplied in the Appendix A.\\

 \noindent {\bf 2.3.} {\em Classical ambiguities and their  relevance on quantum ground.} Let us 
consider the  
  point-splitting
 procedure introduced  in {\bf 1.1} by (\ref{pppro'}).
 The differential operator ${D}_{\mu \nu}(x,y)$ (written in (\ref{Deta}) below putting $\eta=0$ 
therein) is obtained by point-splitting the classical expression for the
 stress-energy tensor (\ref{Tclassic}) \cite{bd,wald94}.   
 The crucial  point is that the classical stress-energy tensor may be replaced by a classically equivalent
 object which, at the quantum level,  breaks such an equivalence. In particular, classically,  we may  
re-define
 \begin{eqnarray}
   T_{\mu\nu}^{(\eta)}(z) \stackrel {\mbox{\scriptsize  def}} {=}    T_{\mu\nu}(z) + \eta\: 
g_{\mu\nu}(z)\: \phi(z) P\phi(z)  \label{Teta}\:,
 \end{eqnarray}
where $\eta\in \bR$ is an arbitrarily fixed pure number and $T_{\mu\nu}(z)$ is given by 
(\ref{Tclassic}).   It is obvious
that     $T_{\mu\nu}^{(\eta)}(z) =    T_{\mu\nu}(z)$  whenever $\phi$ satisfies the field equation 
$P\phi=0$. Therefore,
there is no difference between  the two tensors classically speaking and no ambiguity actually takes 
place through
that way.
On quantum ground things dramatically change since $\langle  \hat{\phi}(x) P\hat{\phi}(x)  
\rangle_\omega\neq 0$,
{\em provided} the left-hand side is defined via point-splitting procedure (see also \cite{hw} where 
the same remark appears in terms of local Wick polynomials). 
 Therefore the harmless
classical ambiguity becomes a true quantum ambiguity.   Actually,  we argue that, without affecting 
the classical
stress-energy tensor, the found ambiguity can be used to clean up the point-splitting procedure.
By this way, the general principle "relevant quantum objects must reduce to corresponding well-
known classical objects in the
{\em formal classical limit}, i.e.,  when quantum observables are replaced by classical observables",  
is preserved.\\
The operator used in the point-splitting procedure corresponding to  $T_{\mu\nu}^{(\eta)}$   is 
obtained
 by means of a point-separation and symmetrization
 of the right-hand side of (\ref{Tclassic}) and (\ref{Teta}). It reads
\begin{eqnarray}
D^{(\eta)}_{(z)\alpha\beta}(x,y) &\stackrel {\mbox{\scriptsize  def}} {=}&  
\frac{1}{2}\left(\delta_\alpha^{\alpha'}(z,x)    \delta_\beta^{\beta'}(z,y)
{\nabla}_{(x)\alpha'}{\nabla}_{(y)\beta'}  + \delta_\alpha^{\alpha'}(z,y)  \delta_\beta^{\beta'}(z,x)
{\nabla}_{(y)\alpha'} {\nabla}_{(x)\beta'} \right)   \nonumber\\
 &-& \frac{1}{2}g_{\alpha\beta}(z)\left(  g^{\gamma\gamma'}(z) 
\delta(z,x)^\mu_{\gamma'}\delta(z,y)^\nu_\gamma {\nabla}_{(x)\mu}  {\nabla}_{(y)\nu}
 +V(z)  \right) \nonumber\\
&+&\xi\left[\left( R_{\alpha\beta}(z) -\frac{1}{2}g_{\alpha\beta}(z) R(z)\right)
 + \frac{g_{\alpha\beta}(z)}{2} \left(\Delta_x   +  \Delta_y \right)   \nonumber \right.   \\
&-& \left. \frac{1}{2}\left(\delta_\alpha^{\alpha'}(z,x) \delta^{\beta'}_{\beta}(z,x)\nabla_{(x)\alpha'}
\nabla_{(x)\beta'} + \delta_\alpha^{\alpha'}(z,y) 
\delta^{\beta'}_\beta(z,y)\nabla_{(y)\alpha'}\nabla_{(y)\beta'}  \right) \right.\nonumber  \\
&+& \left.\eta \:\frac{g_{\alpha\beta}(z)}{2} \left(P_x   +  P_y \right)\right] \:,
\label{Deta}
\end{eqnarray}
$\delta(v,u) $     is the operator of the geodesic transport from $T_uM$ to $T_vM$.
  We aim to show that there is a
choice for $\eta$, $\eta_D$, depending on the dimension of the spacetime manifold $D$ only, such 
that
  \begin{align}
  \langle \hat{T}^{(\eta_D)}_{\mu\nu}(z) \rangle_{\omega} \stackrel {\mbox{\scriptsize  def}} 
{=} \lim_{(x,y) \to (z,z)} {D}^{(\eta_D)}_{(z)\mu \nu}(x,y)[G_\omega^{(1)}(x,y) - Z_{n}(x,y)]
     \label{improved} \:,
 \end{align}
 is physically well behaved.
  To this end a preliminary lemma is necessary.\\

\noindent {\bf 2.4.} {\em A crucial lemma.}
 The following lemma plays a central r\^ole in the proof of Theorem 2.1 concerning the properties of
 the new point-splitting prescription. 
The coefficients of the expansion of $U$ in (\ref{Z2'}), $U_k(z,z)$, which appear below and
$(a|b)$ are defined as in the Appendix A.\\ 

 \noindent {\bf Lemma 2.1.}
 {\em In a smooth $D$-dimensional ($D\geq 2$) spacetime $(M,{\bf g})$
 equipped with the differential operator $P$ in {\em(\ref{operator})},
 the associated Hadamard parametrix {\em(\ref{Z2'}), (\ref{Z2''})} satisfies the following identities,
 where the limits hold uniformly.   \\
{\em {\bf (a)}} If  $n\geq 1$
 \begin{eqnarray}
\lim_{(x,y)\to (z,z)} {P}_x   {Z}_{n}(x,y) =  \lim_{(x,y)\to (z,z)} P_y  {Z}_{n}(x,y) =
  \: \delta_D   \: c_D
   \:  {U}_{D/2}(z,z)\label{lemma}\:.
 \end{eqnarray}
 Above $\delta_D=0$ if $D$ is odd and $\delta_D=1$ if $D$ is even and
 \begin{eqnarray}
 c_D \stackrel {\mbox{\scriptsize  def}} {=} (-1)^{D/2+1} \frac{(2|\frac{D}{2}-1) (D+2)}{2^{D-
1}\pi^{D/2}\Gamma(\frac{D}{2})}
\label{CD}\:,
 \end{eqnarray}
 {\em {\bf (b)}} If  $D$ is even and  $n \geq1$ or $D$ is odd and $n >1$,
 \begin{eqnarray}
\lim_{(x,y)\to (z,z)} {P}_x  \nabla^\mu_{(y)}  {Z}_{n}(x,y) = \lim_{(x,y)\to (z,z)} 
\nabla^\mu_{(x)}
P_y  {Z}_{n}(x,y) =
 \delta_D   k_D
 \nabla^\mu_{(z)} {U}_{D/2}(z,z)\label{lemma'*}
 \end{eqnarray}
 with
  \begin{eqnarray}
 k_D \stackrel {\mbox{\scriptsize  def}} {=} (-1)^{D/2+1} \frac{(2|\frac{D}{2}-
1)D}{2^{D}\pi^{D/2}\Gamma(\frac{D}{2})}
\label{CK}\:.
 \end{eqnarray}
 {\bf (c)} Using the point-splitting prescription to compute $\langle \hat\phi(z) P\hat \phi(z) 
\rangle_{\omega}$ and 
 $\langle P(\hat\phi(z))\hat \phi(z) \rangle_{\omega}$,
\begin{eqnarray}
\langle \hat\phi(z) P\hat \phi(z) \rangle_{\omega}    &\stackrel {\mbox{\scriptsize  def}} {=}& 
\lim_{(x,y)\to (z,z)} P_{(y)}
\left[ G^{(1)}_\omega(x,y)  - Z_{n}(x,y)\right]   =    -\delta_D c_D U_{D/2}(z,z)          
\label{one}\:,   \\
\langle (P\hat\phi(z)) \hat \phi(z) \rangle_{\omega}    &\stackrel {\mbox{\scriptsize  def}} {=}& 
\lim_{(x,y)\to (z,z)} P_{(x)}
\left[ G^{(1)}_\omega(x,y)  - Z_{n}(x,y)\right]   =    -\delta_D c_D U_{D/2}(z,z)          \label{two}  
\:.
\end{eqnarray} 
In particular $\langle \hat\phi(z) P\hat \phi(z) \rangle_{\omega}= \langle P(\hat\phi(z))\hat \phi(z) 
\rangle_{\omega}$.}\\

 \noindent{\em Proof.} See the Appendix B.\\

 \noindent {\em Remark}. With our conventions, when $D$ is even, the anomalous quantum 
correction to the trace
 of the stress-energy tensor  is
  $-2c_DU_{D/2}(z,z)/(D+2)$ \cite{mzp} (and coincides with the
   conformal anomaly if $V\equiv 0$, $\xi=\xi_D$ in (\ref{operator})). Notice that the coefficients 
$U_k(z,z)$ do not depend on 
 either $\omega$ and the scale $\lambda$ used in the definition of $Z_n$. We conclude that
   $\langle \hat\phi(z) P\hat \phi(z) \rangle_{\omega}$ {\bf (i)} does {\em not}
  depend on the scale $\lambda$, {\bf (ii)} does {\em not} depend on $\omega$ and
 {\bf (iii)}  is  proportional to  the anomalous quantum correction to the trace
 of the stress-energy tensor.          \\

 \noindent {\bf 2.5}
 {\em The improved point-splitting procedure.}  Let us show that
 the point-splitting procedure (\ref{improved}) produces a renormalized stress-energy tensor
  which is well behaved and in agreement with
  Wald's four axioms (straightforwardly generalized to the case $V'\not \equiv 0$ when necessary)
  for a particular value of $\eta$ uniquely determined.\\

 \noindent {\bf Theorem 2.1.}
 {\em Let $\omega$ be a  Hadamard  quantum state of a  field $\phi$ on a smooth
globally-hyperbolic $D$-dimensional  ($D\geq 2$) spacetime $(M,{\bf g})$ with  field  operator
{\em(\ref{operator})}. If ${D}^{(\eta)}_{(z)\mu \nu}(x,y)$ is given by   {\em (\ref{Deta})},
consider the symmetric tensor field and the scalar field  locally defined by
\begin{eqnarray}
  z \mapsto     \langle \hat{T}^{(\eta)}_{\mu\nu}(z) \rangle_{\omega} &\stackrel {\mbox{\scriptsize  
def}} {=}&
   \lim_{(x,y) \to (z,z)} {D}^{(\eta)}_{(z)\mu \nu}(x,y)[G_\omega^{(1)}(x,y) - Z_{n}(x,y)] \:,      
\label{improvedD}\\
   z \mapsto  \langle\hat{\phi}^2(z)   \rangle_{\omega,\lambda} &\stackrel {\mbox{\scriptsize  def}} 
{=}&
   \lim_{(x,y) \to (z,z)} [G_\omega^{(1)}(x,y) - Z_{n}(x,y)] \:,      \label{phi2}
\end{eqnarray}
where, respectively, $n\geq 3$ and $n>0$. The following statements hold.\\
{\bf (a)}
 Both $z\mapsto \langle \hat{T}^{(\eta)}_{\mu\nu}(z) \rangle_\omega$ and 
$z \mapsto  \langle\hat{\phi}^2(z)\rangle_\omega$ are smooth and do
not depend on $n$. Moreover, if (and only if) $\eta=\eta_D \stackrel {\mbox{\scriptsize  def}} {=} 
D[2(D+2)]^{-1}$, they
 satisfy
 the analogue of {\em (\ref{conservationT})} for all spacetimes
\begin{eqnarray}
 \nabla^\mu \langle \hat{T}^{(\eta_D)}_{\mu\nu}(z) \rangle_{\omega}  = -\frac{1}{2}   
 \langle \hat{\phi}^2(z) \rangle_{\omega}
 \nabla_\nu  V'(z)\:.
\end{eqnarray}}
{\bf (b)} {\em Concerning the trace of $\langle \hat{T}^{(\eta_D)}_{\mu\nu}(z) \rangle_{\omega}$, 
it
holds
 \begin{eqnarray}
g^{\mu\nu(z)} \langle \hat{T}^{(\eta_D)}_{\mu\nu}(z) \rangle_{\omega}  &=&
 \left[\frac{\xi_D-\xi}{4\xi_D-1}  \Delta  - V(x) \right]  \langle\hat{\phi}^2(z)   \rangle_{\omega}  
\nonumber\\
 &-&\delta_D \frac{2c_D}{D+2} U_{D/2}(z,z) \label{trace'}\:,
\end{eqnarray}
 The  term on the last line,
does not depend on the scale $\lambda>0$ used to define $Z_n$ and  coincides with 
the conformal anomaly for $\xi=\xi_D$, $V\equiv 0$.}\\
{\bf (c)} {\em  If $D$ is even,  $\eta \in \bR$, 
$Q_{\eta,\eta_D}(z) \stackrel {\mbox{\scriptsize  def}} {=} \delta_D (\eta - \eta_D) c_D 
U_{D/2}(z,z)$, it holds
\begin{eqnarray}
 \langle \hat{T}^{(\eta_D)}_{\mu\nu}(z) \rangle_{\omega}   =
 \langle \hat{T}^{(\eta)}_{\mu\nu}(z) \rangle_{\omega}  + g_{\mu\nu}(z) Q_{\eta,\eta_D}(z) \:, 
\label{TQ}
\end{eqnarray}}
 {\bf (d)} {\em Changing the scale $\lambda\to \lambda'>0$  one has, with obvious notation,
 \begin{eqnarray}\langle \hat{T}^{(\eta_D)}_{\mu\nu}(z) \rangle_{\omega,\lambda} -
\langle \hat{T}^{(\eta_D)}_{\mu\nu}(z) \rangle_{\omega,\lambda'} = \delta_D 
\ln\left(\frac{\lambda'}{\lambda}\right) t_{\mu\nu}(z)
\label{difference} \end{eqnarray}
where the  smooth symmetric tensor field $t$  is independent from either the quantum state, 
$\lambda$ and $\lambda'$, 
is conserved for $V'\equiv 0$ and it is built up,
via standard tensor calculus, by employing the metric and the curvature tensors at $z$, $m$,
$\xi$, $V'(z)$ and their covariant derivatives at $z$.}\\
 {\bf (e)} {\em If $(M,{\bf g})$ is the ($D=4$) Minkowski spacetime, $V'\equiv 0$ and $\omega$ 
 is the Minkowski vacuum, there
 is $\lambda>0$ such that
 $\langle \hat{T}^{(\eta_D)}_{\mu\nu}(z) \rangle_{\omega,\lambda}=0$ for all $z\in M$.
 If $m=0$ this holds for every $\lambda>0$.} \\

\noindent {\em Proof.} See the Appendix B.\\

\noindent {\bf Def.2.1 (Quantum averaged stress-energy tensor and field fluctuation).}
{\em  Let $\omega$ be a  Hadamard  quantum state of a  field $\phi$ in a smooth
globally-hyperbolic $D$-dimensional  ($D\geq 2$) spacetime $(M,{\bf g})$ with  field  operator
{\em(\ref{operator})}. Referring
to {\bf Theorem 2.1}, the tensor field
defined in local coordinates by
  $z \mapsto \langle \hat{T}_{\mu\nu}(z) \rangle_{\omega} \stackrel {\mbox{\scriptsize  def}} {=}
\langle \hat{T}^{(\eta_D)}_{\mu\nu}(z) \rangle_{\omega}$   
and the scalar field 
   $z \mapsto  \langle\hat{\phi}^2(z)   \rangle_{\omega}$,
are respectively said  the  {\bf quantum averaged stress-energy tensor} in the state $\omega$
and the {\bf quantum field fluctuation} of the state  $\omega$.}\\

\noindent {\em Remarks.}
{\bf (1)} The point-splitting renormalization defined above turns out to be
in agreement with four Wald's axioms. This can be realized by following the same discussion, 
developed in \cite{wald94}
concerning the standard point-splitting prescription and using the theorem above.\\
{\bf (2)} The need of adding a term to the classical stress-energy tensor
to fulfill the conservation requirement can be heuristically explained as follows.
 As in \cite{bd}, let us assume that there is some functional of the metric corresponding to the one-
loop
effective action:
$$S_\omega[{\bf g}] \stackrel {\mbox{\scriptsize  def}} {=} i\ln \int {\cal D}_{\bf g} \phi \: 
e^{-iS[\phi,{\bf g}]}\:,$$
 where $S$ is the classical action associated with  $P$, and $\omega$ enters the assignment of the
 integration domain.  In this context, the averaged stress-energy tensor is defined as
 $$\langle \hat{T}_{\mu\nu}(z)\rangle_\omega = -\frac{2}{\sqrt{-g(z)}}\frac{\delta S_\omega[{\bf 
g}]}
 {\delta g^{\mu\nu}(z)}\:,$$
 where the functional derivative is evaluated at the actual metric of the spacetime.
The conservation of the left-hand side is equivalent to the (first order) invariance under 
diffeomorphisms of  
$S_\omega[{\bf g}]$. The relevant point is that the measure ${\cal D}_{\bf g} \phi$
  in general must be supposed to depend on the metric  \cite{ha,bd,fu}.
Changing ${\bf g}$ into ${\bf g}'$ by a diffeomorphism, one gets, assuming the
invariance of
$S_\omega[{\bf g}]$ and making explicit the dependence of the measure on the metric
$$0 =  \int {\cal D}_{\bf g} \phi  \left[ \nabla ^\mu \frac{2}{\sqrt{-g(z)}}\frac{\delta J[\phi, {\bf 
g},{\bf g}']}{\delta {g'}^{\mu\nu}(z)}|_{{\bf g}'={\bf g}}
-i \nabla^{\mu} T_{\mu\nu}(z)\right] \: e^{-iS[\phi,{\bf g}]}$$
where $J[\phi, {\bf g},{\bf g}'] {\cal D}_{\bf g}\phi = {\cal D}_{{\bf g}'}\phi$, $J[\phi, {\bf 
g},{\bf g}]=1$.  
The conserved quantity is a term corresponding to the classical stress-energy tensor added to a {\em 
further term} depending on the
functional measure
$$\nabla^\mu \left[ \langle \hat{T}^{(\eta=0)}_{\mu\nu}(z) \rangle_\mu +i\frac{2 e^{iS_\omega[{\bf 
g}]}}{\sqrt{-g(z)}} \int {\cal D}_{\bf g} \phi
\frac{\delta J[\phi, {\bf g},{\bf g}']}{\delta {g'}^{\mu\nu}(z)}|_{{\bf g}'={\bf g}}e^{-iS[\phi,{\bf 
g}]}\right] = 0\:.$$
Therefore the found  term
$\eta_D g_{\mu\nu}(z) \langle \phi(z) P\phi(z) \rangle_\omega$ added  to the classical
stress-energy tensor should be related to the second  term in the brackets above.\\
\noindent {\bf (3)} The functional approach can be implemented via Wick rotation in the case of a 
static
spacetime with compact Cauchy surfaces for finite temperature ($1/\beta $) states and provided $V$ 
does not depend on the
global Killing time. Within that context,   the Euclidean section turns out to be
compact without boundary and $G_\omega^{(1)}$ has to be replaced
with the  unique Green function $G_\beta$, with Euclidean Killing temporal period $\beta$,
of the operator obtained by Wick rotation of $P$.
One expects that the following identity holds
\begin{eqnarray}
-\frac{2}{\sqrt{-g(z)}}\frac{\delta S_{E,\beta}[{\bf g}]}{\delta g^{ab}(z)} = \lim_{(x,y)\to (z,z)}
 {\cal D}^{(\eta_D)}_{(z)ab}(x,y)\left[ G_\beta(x,y) -Z_{n}(x,y) \right]      \label{check}
\end{eqnarray}
where we have replaced  Lorentzian objects by corresponding Euclidean ones and
$a,b$ denote tensor indices in a Euclidean manifold. In a sense, (\ref{check}) can actually be 
rigorously proven as stated in the theorem below.
Indeed, the point-splitting procedure in the right hand side
can be implemented  also in the Euclidean case because 
the parametrices $Z_{\lambda,n}$ ($\lambda$ being the length scale used in the definition 
of the parametrices) 
can be defined also for Euclidean metrics using the same definition given above, 
omitting $\theta(s(x,y))$ in (\ref{Z2''}) and dropping
$|\:\:|$ in the logarithm in  (\ref{Z2'}).
On the other hand, the left-hand side of (\ref{check})
may be interpreted, not depending on the right-hand side, as an Euclidean $\zeta$-function 
regularized stress-energy tensor   $\langle {T}_{ab}(z)\rangle^{(\zeta)}_{\beta,\mu^2}$ 
which naturally introduces an arbitrary mass scale $\mu$  (see \cite{mzp}
where $\sigma(x,y)$ indicates $s(x,y)/2$).  We remind the reader  that, in the same hypotheses,
it is possible to define a $\zeta$-function regularization of the field fluctuation,
$\langle\hat{\phi}^2(z)\rangle^{(\zeta)}_{\beta,\mu^2}$ (see \cite{mzp} and references therein).\\

\noindent {\bf Theorem 2.2.} {\em Let $(M,{\bf g})$ be a  smooth spacetime
endowed with a global Killing time-like vector field  normal to a compact Cauchy surface
and a Klein-Gordon operator $P$ in {\em (\ref{operator})}, where $V'$ does not depend on the 
Killing time.
Consider a compact Euclidean section of the spacetime $(M_\beta, {\bf g}_E)$ obtained by {\em (a)} 
a Wick analytic
continuation with respect to the Killing time and {\em (b)} an identification of the Euclidean time into 
Killing orbits of
period $\beta>0$. Let $G_\beta$ be  the unique (except for null-modes ambiguities) Green function   
of the Euclidean 
Klein-Gordon operator defined on $C^\infty(M_\beta)$ obtained by analytic continuation of $P$. It 
holds
\begin{eqnarray}
\langle {T}_{ab}(z) \rangle^{(\zeta)}_{\beta,\mu^2}  &=&  \langle {T}_{ab}(z)
\rangle_{\beta,\lambda} 
    \label{checked}   \\
 \langle \hat{\phi}^2(z) \rangle^{(\zeta)}_{\beta,\mu^2}  &=&  
\langle {\phi}^2(z) \rangle_{\omega,\lambda}   \label{checked'}
\end{eqnarray}
where $\lambda = c \mu^{-2}$, $c>0$ being some constant and the right-hand sides, and the right-
hand sides of the
{\em (\ref{checked})} and {\em (\ref{checked'})} are defined as in {\em Def.2.1} using $G_\beta$ 
in place of $G^{(1)}_\omega$
and the Euclidean parametrix.}  \\

\noindent {\em Sketch of proof.}
The left-hand side of (\ref{checked}) coincides with
$$\langle \hat{T}^{(\nu_D)}_{ab}(z) \rangle_{\omega, \lambda}  + g_{ab}(z) 
Q_{\nu_D,\eta_D}(z)$$
where $\nu_D = (D-2)/(2D)$, as  shown in Theorem 4.1 of \cite{mzp} provided (using $\hbar=c=1$)
$\lambda$ coincides with $\mu^{-1}$  
with a suitable positive constant factor. (The smooth term  $W$ added to the parametrix which 
appears in the cited theorem 
 can be completely re-absorbed in the logarithmic part of the  parametrix as one can directly show).  
(\ref{TQ}) holds
true also in the Euclidean case as one can trivially show and thus the thesis is proven. 
The proof of (\ref{checked'}) is similar.
   $\Box$

\section{The stress-energy operator in terms of local Wick products.}

 \cite{bfk,bf,hw} contain very significant progress in the definition of perturbative quantum field 
theory 
 in curved spacetime. Those works take advantage from the methods of {\em microlocal analysis} 
\cite{hor1}
 and  the {\em wave front set} characterization of the Hadamard requirement found out by Radzikowski 
\cite{ra}.
  In \cite{bfk} it is proven that,
 in the Fock space generated by a {\em quasifree} Hadamard state, a definition of Wick polynomials 
 (products of field operators evaluated at the same event) can be given with a well-defined
 meaning of operator-valued distributions. That is obtained by the introduction of 
 a normal ordering prescription with respect to a chosen Hadamard state. In the subsequent 
 paper \cite{bf}, it is shown that quantum field theory in curved spacetime gives rise to
  ``ultraviolet divergences'' which are of the same nature as in Minkowski spacetime. 
 This result is achieved by a suitable generalization of the Epstein-Glaser method of renormalization 
 in Minkowski spacetime used to analyze time ordered products of Wick polynomial, involved in the 
perturbative 
 construction of interacting quantum field theory. However the performed analysis shows that 
 quantities which appear at each perturbation order in Minkowski spacetime as renormalized coupling 
 constant are replaced, in curved spacetime, by functions whose dependence upon the spacetime 
points can be 
 arbitrary. In \cite{hw} generalizing the content of \cite{df} and using ideas of \cite{bfk,bf}, 
 it is found that such ambiguity can be reduced to finitely many degrees of
 freedom by imposing a suitable requirement of covariance and locality (which is an appropriate 
replacement
 of the condition of Poincar\'e invariance in Minkowski spacetime).
 The key-step is  a precise notion of {\em local}, {\em covariant quantum field}. In fact, a definition 
 of {\em local} Wick products of field operators in agreement with the given definition of local 
covariant 
 quantum field is stated. Imposing further constrains concerning scaling behavior, appropriate 
continuity 
 properties and commutation relations, two uniqueness theorem are presented
 about local Wick polynomials and their time-ordered products. The only remaining ambiguity 
 consists of a finite number of parameters. Hollands and Wald also sketch a proof of existence 
of local Wick products of 
 field operators in \cite{hw}. The found local Wick products  make use of the Hadamard parametrix 
 only and turn out to be independent from any  preferred Hadamard vacuum state. In principle, by 
means of a straightforward 
 definition to local Wick products of differentiated field, these
 local Wick products may be used to define a well-behaved notion of stress-energy tensor operator.
 However, as remarked in \cite{hw} such a definition would produce a non conserved stress-energy
 tensor. In this section, after a short review of the relevant machinery developed in \cite{hw}, we 
prove 
 how such a problem can be overcome generalizing ideas of Section 2.\\

 \noindent {\bf 3.1.} {\em Normal products and the algebra ${\cal W}(M, {\bf g})$.} 
 From now on, referring to a globally hyperbolic spacetime 
 $(M, {\bf g})$ 
equipped with a Klein-Gordon operator (\ref{operator}), we assume $\mbox{dim}(M)= 4$ and 
$V'\equiv 0$ 
 in (\ref{operator}). In the following, for $n=1,2,\ldots$,  ${\cal D}(M^n)$ denotes the space of 
smooth 
 compactly-supported complex 
 functions on $M^n$
 and ${\cal D}_n(M) \subset {\cal D}(M^n)$ indicates the subspace containing  the functions  
 which are symmetric under interchange of every pair of arguments. 

 In the remaining part of the work we make use of some mathematical tools defined in microlocal 
analysis.
  (See chapter VIII of \cite{hor1} concerning the notion of wave front set and \cite{ra} concerning
 the microlocal analysis characterization of the Hadamard requirement.)      
 Preserving the usual seminorm-induced topology on ${\cal D}(M)$,
 all definitions and theorems about distributions $u\in {\cal D}'(M)$  (chapter VI of \cite{hor1})
 can straightforwardly be re-stated for {\em vector-valued} distributions and in turn, partially, for 
{\em operator-valued} distributions
 on ${\cal D}(M)$. That is, respectively, continuous linear maps $v: {\cal D}(M) \to {\cal H}$, 
${\cal H}$ being
  a Hilbert space, and
 continuous  linear maps $A: {\cal D}(M) \to {\cal A}$, 
${\cal A}$ being a space of operators on ${\cal H}$ (with common domain) endowed with the {\em 
strong} Hilbert-space topology.
The content of Chapter VIII of \cite{hor1} may straightforwardly be generalized to vector-valued 
distributions. 
 
 In this part we consider {\em quasifree} \cite{kw,wald94} states $\omega$.  
 Referring to {\bf 2.2}, this means that the $n$-point functions 
are obtained by functionally differentiating with respect to $f$ the {\em formal} identity
$$\omega(e^{i\phi(f)}) = e^{-\frac{1}{2}\omega(\phi(f)\phi(f))}\:.$$
In that case there is a GNS Hilbert space ${\cal H}_\omega$ which is a bosonic Fock space, 
$\Omega_\omega\in {\cal H}_\omega$ is the  vacuum vector therein, operators $\hat \phi(f)$ are essentially 
self-adjoint
on ${\cal D}_\omega$ if $f\in {\cal D}(M)$ is real and Weyl's  relations are fulfilled
by the one-parameter groups generated by operators $\overline{\hat \phi(f)}$.

 Let us introduce normal Wick products {\em defined with respect to a reference quasifree Hadamard 
state $\omega$} \cite{bf,hw}. 
 Fix a GNS triple for $\omega$, $({\cal H}_\omega,\Pi_\omega,\Omega_\omega)$
 and consider the algebra of operators with domain ${\cal D}_\omega$, ${\cal A}_\omega(M, {\bf 
g})$ (see {\bf 2.2}).
 {\em From now on, we write $\hat \phi$ instead of $\hat \phi_\omega$ whenever it does not give rise 
to misunderstandings.}
  For $n\geq 1$, define the symmetric operator-valued linear map, $\hat{W}_{\omega, n} : {\cal 
D}_n(M) \to {\cal A}_\omega(M, {\bf g})$,
  given by the formal symmetric kernel 
 $$\hat{W}_{\omega, n}(x_1,\ldots,x_n) \stackrel {\mbox{\scriptsize  def}} {=} 
:\spa\hat\phi(x_1)\cdots \hat\phi(x_n)\colon\sp_{\omega}$$
 \begin{align}
  \stackrel {\mbox{\scriptsize  def}} {=}  \left(\prod_{j=1}^n \frac{1}{i (-g(x_j))^{1/2}} 
\right)\left.\frac{\delta^n e^{\left\{ \frac{1}{2}
  \int \int \omega(x,y)f(x)f(y) d\mu_{\bf g}(x)d\mu_{\bf g}(y)+ i \int \hat\phi(z) f(z) d\mu_{\bf 
g}(z)\right\}}  }{\delta f(x_1)\cdots \delta f(x_n)}  \right \vert_{f\equiv 0}
  \label{normal}
 \end{align}
 where the result of the formal functional derivative is supposed to be {\em symmetrized}, and thus
 only the symmetric part of $\omega$, i.e., $G_\omega^{(1)}$, takes place in in (\ref{normal}).
  $\hat{\phi}(x)$ is the formal kernel of 
 $\hat\phi (= \hat\phi_\omega)$, $\omega(x,y)$ is the formal kernel of $\omega$. Finally define
 $\hat{W}_{\omega, 0} \stackrel {\mbox{\scriptsize  def}} {=} I$ the unit of ${\cal A}_\omega(M, 
{\bf g})$.\\
  The operators $\hat W_{\omega, n}(h)$ can be extended (or directly defined) \cite{bf,hw} to a 
dense
  invariant subspace of ${\cal H}_\omega$, the ``microlocal domain of 
  smoothness'' \cite{hw}, 
  $D_\omega \supset {\cal D}_\omega$,
  which is contained in the self-adjoint extension of every operator $\hat\phi(f)$
  smeared by {\em real} $f\in {\cal D}(M)$\footnote{Therefore, Weyl's commutation relations, and 
thus
  bosonic  commutation relations on $D_\omega$, are preserved.}.\\
  {\em From now on we  assume that every considered operator is defined on $D_\omega$.}\\
  $D_\omega$  enjoys two relevant properties. (a) Every map $h\mapsto \hat W_{\omega, n}(h)$, 
$h\in {\cal D}_n(M)$,  
  defines a symmetric operator-valued {\em distribution}. (b) Those operator-valued distributions 
  may give rise to operators which can be interpreted as {\em products of field operators evaluated at 
the same event}. 
  This is because every $\hat{W}_{\omega, n}$
  can be smeared by a suitable class of {\em distributions} and, in particular, $\hat{W}_{\omega, 
n}(f\delta_n)$
  can be interpreted as $:\spa\hat\phi^n(f)\colon \sp_\omega$ if $f\in {\cal D}(M)$ and $\delta_n$ is 
the  
  distribution:
  $$\int_M h(x_1,\ldots,x_n) \delta_n(x_1,\ldots,x_n) d\mu_{\bf g}(x_1) \cdots d\mu_{\bf g}(x_n)        
\stackrel {\mbox{\scriptsize  def}} {=}
 \int_M h(x,x,\ldots,x) d\mu_{\bf g}(x)\:.$$
  Let us summarize the proof of this remarkable result following \cite{hw}.
  By Lemma 2.2 in \cite{bf}, if $\Psi\in D_\omega$ the wave front set of the vector-valued 
distributions 
  $t\mapsto \hat{W}_{\omega, n}(t) \Psi$,
  $WF\left(\hat{W}_{\omega, n}(\cdot) \Psi\right)$ \cite{hor1}, is contained in the set 
  \begin{eqnarray}
  {\bf F}_n(M,{\bf g}) \stackrel {\mbox{\scriptsize  def}} {=} \{(x_1,k_1,\ldots, x_n,k_n) \in 
(T^*M)^n \setminus \{0\}
  | k_i\in V^-_{x_i}, i=1,\ldots,n\}  \label{Fn}\:,
  \end{eqnarray}
  $V_x^{+/-}$ denoting the set of all nonzero time-like and null
   co-vectors at $x$ which are future/past directed.  Theorem 8.2.10 in \cite{hor1} states that
   if the wave front sets of two distributions $u,v \in {\cal D}'(N)$, $N$ being any manifold, satisfy 
$WF(u)+ WF(v)\not \ni \{0\}$,
   then a pointwise product between $u$ and $v$, $u\odot v$ can be unambiguously defined  giving 
rise to a distribution
   of ${\cal D}'(N)$. The theorem can be straightforwardly generalized to vector-valued distributions.
   In our case we are allowed to define the product between a  distribution $t$
   and a vector-valued distribution $\hat{W}_{\omega, n}(\cdot) \Psi$ 
   provided $WF(t)+ {\bf F}_n(M,{\bf g})\not \ni \{0\}$. To this end
   define
   $${\cal E}'_n(M, {\bf g}) \stackrel {\mbox{\scriptsize  def}} {=}  \left\{ t \in {\cal D}_n'(M)\:|\:
   \mbox{$supp\:\: t$ is compact, $WF(t) \subset {\bf G}_n(M,{\bf g})$ } \right\}$$
   where
   $${\bf G}_n(M,{\bf g}) \stackrel {\mbox{\scriptsize  def}} {=} T^*M^n\setminus
   \left( \bigcup_{x\in M}(V_x^+)^n \cup \bigcup_{x\in M}(V_x^-)^n\right)\:.$$
   It holds $WF(t)+ {\bf F}_n(M,{\bf g})\not \ni \{0\}$ for $t\in {\cal E}'_n(M, {\bf g})$. By 
consequence the product, 
   $t \odot  \hat W_{\omega, n}\Psi$,
   of the distributions $t$ and $\hat{W}_{\omega, n}(\cdot) \Psi$ can be defined for
   every $\Psi \in D_\omega$ and it is possible to show that $\left(t \odot  \hat W_{\omega, 
n}\Psi\right)(f) \in D_\omega$
   for every $f\in {\cal D}_n(M)$.
    In turn, varying $\Psi \in D_\omega$, one straightforwardly gets a well-defined  
    {\em operator-valued}  distribution $t \odot  \hat W_{\omega, n}$. \\
    Summarizing: if $t\in {\cal E}'_n(M, {\bf g})$, $n\in \bN$, it is well-defined an 
    operator-valued symmetric distribution
     ${\cal D}_n(M) \ni f \mapsto \left(t \odot  \hat W_{\omega, n}\right)(f)$,
     with  values defined in the dense invariant domain $D_\omega$. \\
    To conclude we notice that if $t\in {\cal E}'_n(M,{\bf g})$, $\hat W_{\omega, n}$ 
    can  be {\em smeared} by $t$ making use of the
   following definition.  Since, for all $\Psi \in D_\omega$,
   $supp\: ( t\odot \hat{W}_{\omega,n}\Psi) \subset supp\: t$\footnote{It can be shown using the 
continuity of the product with respect to the
   H\"ormander pseudo topology and  theorem 6.2.3 of \cite{hor1} which assures that each 
distribution is the limit in that pseudo topology
   of a sequence of smooth functions and the fact that the convergence in the pseudo topology implies 
the usual convergence
   in ${\cal D}'$.},  take  $f\in {\cal D}_n(M)$ such that $f(x_1,\ldots,x_{n})=1$ for
   $(x_1,\ldots,x_n)\in supp\: t$  and define the operator, with domain $D_\omega$,
   $$ \hat W_{\omega, n}(t) \stackrel {\mbox{\scriptsize  def}} {=}  \left(t \odot \hat 
W_{\omega,n}\right)(f)\:.$$
   It is simply proven that the definition does not depend on the used $f$
    and the new smearing operation reduces with the usual one for  $t\in {\cal D}_n(M) \subset {\cal 
E}'_n(M, {\bf g})$. Finally,
   since $f\delta_n \in {\cal E}'_n(M, {\bf g})$ if $f\in {\cal D}(M)$, the following operator-valued 
distribution 
   is well-defined
   on  $D_\omega$,
   $$f\mapsto \:\: :\spa\hat\phi^n(f)\colon \sp_\omega
   \stackrel {\mbox{\scriptsize  def}} {=}\hat{W}_{\omega,n}(f\delta_n)\:,$$
   $ :\spa\hat\phi^n(f)\colon \sp_\omega$ is called
   {\em normal ordered product of $n$ field operators with respect to $\omega$}. 
    Generalized normal ordered Wick products of $k$ fields, $\:\: :\spa\hat\phi^{n_1}(f_1) \cdots 
\hat\phi^{n_k}(f_k)\colon \sp_\omega$
    are similarly defined  \cite{hw}.

  Given a quasifree Hadamard state $\omega$ and a GNS representation,
  ${\cal W}_\omega(M,{\bf g})$ is the $*$-algebra generated by
  $I$ and the operators $\hat{W}_{\omega,n}(t)$
  for all $n\in \bN$ and $t\in {\cal E}'_n(M,{\bf g})$ with involution given by 
   $\hat{W}_{\omega,n}(t)^* \stackrel {\mbox{\scriptsize  def}} {=}
  \hat{W}_{\omega,n}(t)^\dagger\spa\rest_{D_\omega} (= \hat{W}_{\omega,n}(\overline{t}))$.
  ${\cal A}_\omega(M,{\bf g})$ 
  turns out to be a sub $*$-algebra of ${\cal W}_\omega(M,{\bf g})$ since one finds  that
  $\hat \phi_\omega(f) = \:\: :\spa \hat \phi(f)\colon\sp_\omega$ for $f\in {\cal D}(M)$.
   
   Different GNS triples for the same $\omega$ give rise to unitary equivalent  
   algebras ${\cal W}_\omega(M,{\bf g})$ by GNS's theorem. However,
   if  $\omega,\omega'$ are two quasifree Hadamard states, 
   ${\cal W}_\omega(M,{\bf g}), {\cal W}_{\omega'}(M,{\bf g})$ 
 are isomorphic  (not unitary
 in general) under a canonical 
 $*$-isomorphism $\alpha_{\omega' \omega} : {\cal W}_\omega(M,{\bf g}) \to  {\cal 
W}_{\omega'}(M,{\bf g})$,
 as shown in Lemma 2.1 in \cite{hw}.
  These $*$-isomorphisms also satisfy, $\alpha_{\omega'' \omega'}\circ \alpha_{\omega' \omega} =
  \alpha_{\omega'' \omega} $ and  $\alpha_{\omega' \omega} (\hat \phi_\omega(t)) = \:\: \hat 
\phi_{\omega'}(t)$,
 but in general, for $n>1$, $\alpha_{\omega' \omega} ( :\spa \hat \phi^n(t) \colon\sp _{\omega} ) 
\neq\:\:
 :\spa \hat \phi^n(t) \colon\sp _{\omega'} $.

One can define  
an abstract $*$-algebra ${\cal W}(M,{\bf g})$, isomorphic to each $*$-algebra  ${\cal 
W}_\omega(M,{\bf g})$ by
$*$-isomorphisms  $\alpha_\omega : {\cal W}(M,{\bf g}) \to {\cal W}_\omega(M,{\bf g})$ such 
that,  if $\omega,\omega'$ are quasifree 
Hadamard  states, 
$\alpha_{\omega'}\circ \alpha_\omega^{-1} = \alpha_{\omega' \omega}$. 
As above ${\cal A}(M,{\bf g})$ is $*$-isomorphic to a sub $*$-algebra of ${\cal W}(M,{\bf g})$ 
and
$\alpha_\omega( \phi(t) ) =\:\: :\spa\hat\phi(t)\colon\sp_\omega$. Elements $W_{\omega,n}(t)$ and 
$:\spa\phi^n(f)\colon\sp_\omega$ are defined in ${\cal W}(M,{\bf g})$ via (\ref{normal}).\\

 \noindent {\bf 3.2.} {\em Local Wick products.} Following \cite{hw}, a {\bf quantum field in one 
variable}  $\Phi$
 is an assignment which associates with every globally hyperbolic spacetime $(M,{\bf g})$
 a distribution $\Phi[{\bf g}]$ taking values in the algebra ${\cal W}(M,{\bf g})$. $\Phi$, is said 
{\bf local and covariant} \cite{hw}
 if it satisfies the  following  \\
 {\bf Locality and Covariance requirement}: {\em For any embedding $\chi$ from a spacetime $(N, 
{\bf g}')$ 
 into another spacetime $(M, {\bf g})$ which is isometric (thus ${\bf g}'= \chi^* {\bf g}$ ) and 
causally preserving\footnote{That is $\chi$
 preserves the time orientation and  $J^+(p)\cap J^-(q) \subset \chi(N)$ if  $p,q\in \chi(N)$.},  it 
holds\\
 $${i}_\chi(\Phi[{\bf g}'](f)) = \Phi[{\bf g}](f\circ \chi^{-1})\:\:\:\:\:\: \mbox{for all $f \in {\cal 
D}(N)$}\:.$$
 Above ${i}_\chi : {\cal W}(N, {\bf g}') \to {\cal W}(M, {\bf g})$ is the  injective 
 $*$-algebra homomorphism such that if $\omega$ is a quasifree Hadamard state on $(M,{\bf g})$ 
and 
 $\omega'(x,y) = \omega(\chi(x),\chi(y))$, we have, 
\begin{eqnarray} 
i_\chi({W}_{\omega',n}(t)) = {W}_{\omega,n}(t\circ \chi^{-1}_n) \:\:\:\:\:\: 
 \mbox{for all $n\in \bN$, $t \in {\cal E}'_n(N, {\bf g}')$}\:.
\label{chi} \end{eqnarray}
 where $\chi^{-1}$ is defined on $\chi(N)$} and $(t\circ \chi^{-1}_n)(x_1,\ldots,x_n) \stackrel 
{\mbox{\scriptsize  def}} {=} 
 t(\chi^{-1}(x_1),\ldots, \chi^{-1}(x_n))$. 
 {\em The generalization to (locally and covariant) quantum field in $n$-variables  is 
straightforward.}\\
   It is worth stressing that the notion of local covariant field is not trivial. For instance,
   any assignment of the form $(M,{\bf g})\mapsto \omega(M,{\bf g})$
   where $\omega(M,{\bf g})$ are quasifree Hadamard states, does {\em not} define a local covariant 
quantum field
   by the map $(M,{\bf g})\mapsto \: :{\phi}^2\colon\spa_{\omega(M,{\bf g})}$ \cite{hw}.
   
 In \cite{hw}, Hollands and Wald sketched a proof of existence of local and covariant quantum fields 
in terms
 of {\em local  Wick products} of field operators.  Let us review the construction of these Wick 
products 
  also making some  technical improvements.\\
 As $M$ is strongly causal \cite{BEE,wald84}, there is a topological base of open sets
 $N$ such that each $N$ is contained in a convex normal neighborhood,
  each inclusion map $i: N\to M$ is causally preserving and each $N$
is  globally hyperbolic with respect to the induced metric. 
 We call {\em causal domains} these open neighborhoods $N$. \\
 Let $N\subset M$ be a causal domain.
 The main idea to built up local Wick products \cite{hw} consists of a suitable use 
 of the Hadamard parametrix which is
  locally and covariantly defined in the globally hyperbolic spacetime $(N, {\bf g}\spa \rest_N)$ in 
terms of the metric \cite{hw}.
 In fact, it is possible to define a suitable distribution $H\in {\cal D}'(N\times N)$  such that, 
 every distribution $H - Z_n\spa\rest_{N\times N}$ is a function of $(x,y) \in N\times N$ 
which is 
 smooth for $x\neq y$ and  
 with vanishing derivatives for $x=y$ up to the order $n$. 
 Then define the elements of ${\cal W}(N,{\bf g}\spa\rest_{N})$, $W_{H,0} \stackrel 
{\mbox{\scriptsize  def}} {=}1$ and
$W_{H,n}$ given 
 by  (\ref{normal}) with $\omega$ replaced by $H$ and $\hat\phi$ replaced by $\phi\in {\cal 
W}(N,{\bf g}\spa\rest_{N})$.
 These distributions enjoy the same smoothness properties of 
 $\hat W_{\omega,n}$
 for every  quasifree Hadamard state $\omega$ because 
  $G^{(1)}_{\omega}\spa\rest_{N\times N} - H = 
  (G^{(1)}_{\omega} - Z_n)\spa\rest_{N\times N} - (H - Z_n\spa\rest_{N\times N})$ 
  is smooth on $N\times N$ and all of its derivative (of any order) must vanish at $x=y$  since
  $n$ is arbitrary.
  In particular every $W_{H,n}$ can be smeared by distributions of ${\cal E}'_n(N, {\bf 
g}\spa\rest_N)$.
  The {\em local Wick products} (on $N$) found by Hollands and Wald in \cite{hw} are the 
elements of
  ${\cal W}(N,{\bf g}\spa\rest_N)$ of the form, with $f \in {\cal D}(N)$,
  $$:\phi^n(f)\colon\spa_H \stackrel {\mbox{\scriptsize  def}} {=} W_{H,n}(f\delta_n)\:,$$  
  
  A few words on the construction of $H$ are necessary. $H$ is given as follows
  \begin{eqnarray}
H \stackrel {\mbox{\scriptsize  def}} {=} Re (H^{(+)})    \label{H}\:.
\end{eqnarray}
Using  definitions and notation as in the Appendix A, the distribution
 $H^{(+)} \in {\cal D}'(N\times N)$ is defined, in the sense of the $\epsilon$-prescription,  by a 
re-arrangement
 of  the kernel of $Z^{(+)}_{n}$ (\ref{W2'}) with $D=4$, i.e.,
 \begin{eqnarray}
\beta^{(1)}_4 \: \frac{{U}(x,y) }{s_{\epsilon,T}(x,y)}+ \beta^{(2)}_4 {V}^{(\infty)}(x,y) \: 
\ln
 \frac{s_{\epsilon,T}(x,y)}{\lambda^2}  \label{Z}\:.
 \end{eqnarray}
 (Similarly to $Z^{(+)}_{n}$, $H^{(+)}$ does not depend on the choice of the temporal 
 coordinate $T$.)
Above
  $$V^{(\infty)}(x,y) \stackrel {\mbox{\scriptsize  def}} {=} \sum_{k=0}^{+\infty} \frac{1}{2^{k-
1} k!} \:U_{k+1}(x,y)\: 
  \psi\left(\frac{s(x,y)}{\alpha_k}\right) \:s^k(x,y)\:. $$
$\psi: \bR \to \bR$ is some smooth map with $\psi(x)=1$  for $|x|<1/2$ and $\psi(x)=0$ for $|x|>1$ 
and
 $\alpha_k >0$ for all $k\in \bN$.
 The series above converges to a smooth function which 
 vanishes with all of its derivatives at $x=y$, provided  the reals $\alpha_k$'s tend to zero sufficiently
 fast (see \cite{gunther}).

From now on we omit the restriction symbol $\rest_N$ and $\rest_{N\times N}$ whenever these
 are implicit in the context. Our aim to extend the given definitions to the whole 
manifold $M$
(and not only $N$) and generalize to differentiated  field the notion of 
local Wick products. We have a preliminary proposition.\\

 \noindent {\bf Proposition 3.1.} {\em Referring to the given definitions, the sub $*$-algebra of  
${\cal W}(N,{\bf g})$, ${\cal W}_H(N,{\bf g})$, 
 generated by $W_{H,n}(t)$,  $t\in {\cal E}'_n(N,{\bf g})$,  $n=0,1,\ldots$. {\bf (a)} 
${\cal W}_H(N,{\bf g})$ coincides with ${\cal W}(N,{\bf g})$ it-self and
 {\bf (b)} is naturally $*$-isomorphic  to the sub $*$-algebra of ${\cal W}(M,{\bf g})$ whose 
elements are smeared by
distributions with support in $N^n$. In this sense $W_{H,n}(t)\in {\cal W}(M,{\bf g})$, $n\in 
\bN$.} \\

 \noindent {\em Proof.} 
 (a)  Fix a Hadamard state $\omega$ in
  $(N, {\bf g})$ and generate ${\cal W}(N,{\bf g})$ by elements $W_{\omega,n}(t)$.
  Define the $*$-isomorphism $\alpha:{\cal W}_H(N,{\bf g}) \to {\cal W}(N,{\bf g})$  
  as in the proof of Lemma 2.1 in \cite{hw} with $d \stackrel {\mbox{\scriptsize  def}} {=} 
G^{(1)}_\omega - H$.
 (The reality of $d$ is assured by the fact that  $H$ is real.) 
$\alpha$, in fact, is the identity map in ${\cal W}_H(N,{\bf g})$.
 (b) It is a direct consequence of the existence of the  natural injective $*$-homomorphism defined in 
Lemma 3.1 in \cite{hw}. 
 $\Box$\\

 \noindent In order to define local Wick products of field operators, consider $n$ linear differential 
operators $K_i$, acting
 on functions of ${\cal D}(M)$,  with the form
    \begin{eqnarray}
 K_i \stackrel {\mbox{\scriptsize  def}} {=}   a_{(i0)} + \nabla_{a_{(i1)}} + \nabla^2_{a_{(i2)}} + 
 \cdots +\nabla^L_{a_{(iL_i)}}\:,\label{opdif}
 \end{eqnarray}
  where
  $a_{(i0)}\in C^\infty(M; \bC)$ and, for $k>0$,  $a_{(ik)}$ is a smooth complex contravariant  
tensor field of order $k$ defined on $M$.
 $\nabla^k_{a} :{\cal D}(M) \to {\cal D}(M)$ is defined, in each local chart, by
 $$\nabla^k_{a(x)} = a^{\mu_1\ldots \mu_k}(x) \nabla_{\mu_1(x)}\cdots  \nabla_{\mu_k(x)}\:.$$
 $t_n[K_1,\ldots, K_n,f]\in {\cal D}'(M^n)$ is the compactly supported in $N^n$ distribution with 
formal kernel
   \begin{eqnarray}
t_n[K_1,\ldots, K_n,f](x_1,\ldots,x_n) \stackrel {\mbox{\scriptsize  def}} {=}
{^tK_n}_{(x_n)} \:{^tK_{n-1}}_{(x_{n-1})}\ldots  \: {^tK_{1}}_{(x_1)}
 f(x_1)   \delta_n(x_1,\ldots x_n)    
\label{tfdev}\:,
 \end{eqnarray}
 where {\em the right-hand side is supposed to be symmetrized} in $x_1,\ldots,x_n$. Above
 $f\in {\cal D}(N)$,  ${\cal D}(N)$ being  identified  with the  subspace of ${\cal D}(M)$ containing 
the functions with support in $N$.  The transposed
 operator  ${^tK_i}$ is defined as usual by "covariant" integration by parts with respect to ${K_i}$ 
\cite{friedlander}.
 As a general result, $WF(\partial u)\subset WF(u)$ and $WF(hu)\subset WF(u)$ if $h$ is smooth.
 By consequence, for every $f\in {\cal D}(N)$ and operators $K_i$, $t_n[K_1,\ldots, K_n,f]\in {\cal 
E}'_n(M, {\bf g})$
 because
 $WF(t_n[K_1,\ldots, K_n,f]) \subset WF(f\delta_n)
 \subset \{(x_1,k_1;\ldots ;x_n,k_n)\in T^*M^n\setminus \{0\}\:|\: \sum_i k_i =0\}$ which is 
 a subset of ${\bf G}_n(M,{\bf g})$.
 This result enables us to state the following definition.  \\

\noindent  {\bf Def 3.1 (Local wick products of (differentiated) fields I).}
{\em Let  $N$ be a causal domain in a globally hyperbolic spacetime $M$ with $H$  defined  in {\em 
(\ref{H})}.
 The {\bf local Wick product of} $n$ ({\bf differentiated fields})
 {\bf generated by $n$ operators $K_i$ and $f\in {\cal D}(M)$ with $supp\:\: f \subset N$} is
   \begin{eqnarray}
:\spa K_1\phi \cdots K_n\phi (f)\colon\sp_H   
\stackrel {\mbox{\scriptsize  def}} {=} W_{H,n}(t_n[K_1,\ldots, K_n,f]) \in {\cal W}(M, {\bf 
g})\:.          
 \label{phind}
 \end{eqnarray}}
   The definition can be improved dropping  the  restriction $supp\:f \subset N$ as follows. 
A preliminary lemma is necessary.   \\

 \noindent  {\bf Lemma 3.1.} {\em Referring to {\bf Def 3.1}, the following statements hold.\\
{\bf (a)} The local Wick products on a causal domain $N\subset M$,
 do not depend on the arbitrary terms $\psi$ and $\{\alpha_k\}$ used in the definition of $H$
 (but may depend on the length scale $\lambda$).\\
 {\bf (b)} If $N'\subset M$ is another causal domain with $N\cap N' \neq \emptyset$ and 
   $:\spa K_1\phi \cdots K_n\phi (f)\colon\sp_{H'}$ denote a local Wick product of differentiated 
fields operators
  defined on $N'$ using the same length scale $\lambda$ as in $N$, then
  \begin{eqnarray}
   :\spa K_1\phi \cdots K_n\phi (f) \colon\sp_H =\:\: :\spa K_1\phi \cdots 
K_n\phi(f)\colon\sp_{H'}\label{compat1}
   \end{eqnarray}
    and for any choice of operators $K_i$ and $f\in {\cal D}(N\cap N')$.}\\
    
  \noindent{\em Proof.} See the Appendix B.\\

   \noindent {\bf Def. 3.2. (Local Wick products of (differentiated) fields II)} {\em Referring
    to {\bf Def. 3.1}, consider an open cover  $\{N_i\}$ of $M$ made of causal domains with
    distributions $H_i$ defined with the same scale length $\lambda$.
    Take a smooth partition of the unity $\{\chi_{i_j}\}$, 
    with $supp\:\: \chi_{i_j} \subset O_{i_j}\subset N_i$, $\{O_{i_j}\}$ being a locally finite
    refinement of $\{ N_i\}$.  The {\bf local Wick product of } $n$ ({\bf differentiated}) {\bf fields
    generated by $n$ operators  $K_i$ and} $f \in {\cal D}(M)$
    is the element  of  ${\cal W}(M, {\bf g})$
    \begin{eqnarray}
     :\spa K_1\phi \cdots K_n \phi (f) \colon\sp   \stackrel {\mbox{\scriptsize  def}} {=}
    \sum_{i,j} :\spa K_1\phi \cdots K_n\phi (\chi_{i_j} f)\colon\sp_{H_{i}}   \:, \label{defgen}   
     \end{eqnarray}}
   {\em Remark.} Only a finite number of non vanishing terms are summed in the right-hand side of 
(\ref{defgen})
    as a consequence of the locally finiteness of the cover $\{O_{i_j}\}$ and the compactness of $supp \:f$.
    Moreover, by (a) of Lemma 3.1 and the linearity on ${\cal E}_n'(M, {\bf g})$ of the involved 
    distributions,  the given definition does not depend 
     on the functions $\psi$ and constants $\{\alpha_k\}$ used in the definition of $H$. 
    By (b) of Lemma 3.1 the definition is independent from  
    the choice of the cover and on the  partition of the unity.\\

    \noindent The (differentiated) local Wick products enjoy the following properties.  \\

    \noindent {\bf Proposition 3.2.} {\em Let $(M, {\bf g})$ be  a four-dimensional globally 
hyperbolic spacetime 
    with a Klein-Gordon
  operator {\em (\ref{operator})} with $V'\equiv 0$.
  Given $n>0$ operators $K_i$, the following statements hold.\\
          {\bf (a)} Given $a,b \in \bC$, $f,h\in {\cal D}(M)$
	  \begin{eqnarray}
	  :\spa \phi(f)\colon &=& \phi(f)\:,\\
	  :\spa K_1\phi \cdots K_n\phi(f) \colon\sp^* &=&
  \:\overline{K_1}\phi \cdots \overline{K_n}\phi (\overline{f})\colon  \:,\\
  :\spa K_1\phi \cdots K_n\phi (a f+b h) \colon  &=&
  a:\spa K_1\phi \cdots K_n\phi (f) \colon 
  + b :\spa K_1\phi \cdots K_n\phi(h)\colon \:.
  \end{eqnarray}
     {\bf (b)} If $\omega$ is a quasifree Hadamard state  on $M$, define 
     $:\spa K_1\hat\phi_{\omega} \cdots K_n\hat\phi_{\omega} (f) \colon\spa \in {\cal 
W}_\omega(M,{\bf g})$
     with
     $f\in {\cal D}(M)$,
     by {\bf Def 3.2} using the operators $\hat \phi_\omega(h)$
     of a  GNS representation of $\omega$. It holds
        \begin{eqnarray}
  :\spa K_1\hat\phi_{\omega} \cdots K_n\hat\phi_{\omega} (f) \colon   =
    \alpha_\omega(:\spa K_1\phi\cdots K_n\phi(f)\colon\spa) \:
     \label{compat2} \end{eqnarray}
      Moreover, varying $f\in {\cal D}(M)$, the left-hand side gives rise to an operator-valued 
distribution
      $f \mapsto :\spa K_1\hat\phi_{\omega} \cdots K_n\hat\phi_{\omega}(f) \colon $
      defined on the dense invariant subspace $D_\omega$.\\
     {\bf (c)} For $f\in {\cal D}(M)$
     \begin{eqnarray}:\spa\overline{K_1}\hat\phi_{\omega} \cdots \overline{K_n}\hat\phi_{\omega} 
(\overline{f})\colon 
  \subset \:\::\spa K_1\hat\phi_{\omega} \cdots K_n\hat\phi_{\omega} 
(f)\colon\sp^\dagger\end{eqnarray}
     {\bf (d)} If $\omega,\omega'$ are Hadamard  states on
  $M$ and $f\in {\cal D}(M)$,
 \begin{eqnarray}\alpha_{\omega,\omega'}(:\spa K_1\hat\phi_{\omega} \cdots 
K_n\hat\phi_{\omega} (f)\colon)=
   \:\::\spa K_1\hat\phi_{\omega'} \cdots K_n\hat\phi_{\omega'}(f) \colon\:.\end{eqnarray}}

 \noindent {\em Remark.} (d) does not hold for normal products defined w.r.t. any quasifree 
Hadamard state 
 $\omega$. \\

  \noindent {\em Sketch of proof.}  (a) is  direct consequences of the given definitions, the reality of 
$H$
  and the linearity of all the involved distributions on ${\cal E}_n'(M, {\bf g})$.
 (b) The continuity with respect to the strong 
Hilbert-space topology is the only non trivial point. It can  be shown as 
follows. Take a sequence of
  functions $\{f_j\}\subset {\cal D}(M)$ with $f_j \to f$ in ${\cal D}(M)$. In particular, this implies 
that there is a 
  compact $K$
  with $supp \:\: f_j, supp\:\: f \subset K$ for $j>j_0$.  By Def.3.2, it is sufficient 
  to prove the continuity when the supports of
  test functions belong to a common causal domain $N \subset M$, i.e.,  
   $\alpha_\omega(W_{H,n}(t_n[K_1,\ldots, K_n, f_j ]))\Psi
  \to \alpha_\omega(W_{H,n}(t_n[K_1,\ldots, K_n^, f]))\Psi$ if     $f_k \to f$ in ${\cal D}(N)$
and $\Psi \in D_\omega$.
  By the definition of ${\cal W}(M, {\bf g})$ and  the isomorphism $\alpha_\omega$ (see {\bf 3.1}),
  it is sufficient to show that $t_n[K_1,\ldots, K_n, f_j] \to t_n[K_1,\ldots, K_n, f]$
  in the closed conic set $\Gamma_n = \{(x_1,k_1;\ldots ;x_n,k_n)\in T^*M^n\setminus \{0\}\:|\: 
\sum_i k_i =0\}$ 
  which contains the wave front set of all involved distributions,
  if $f_j \to f$ in
  ${\cal D}(N)$. The proof of the required convergence property is quite technical and it  is proven
  in the Appendix B. (c) is a trivial consequence of the fact that $\alpha_\omega$ is a $*$-
isomorphism and the definition
  of the involution on ${\cal W}(M, {\bf g})$.  (d) is a trivial consequence of (\ref{compat2}) and the 
identity
  $\alpha_{\omega,\omega'} = \alpha_{\omega'} \circ \alpha_\omega^{-1}$.
  $\Box$\\
  
  We can state a generalized locality and covariance requirement. A {\bf differentiated quantum field in 
one variable}  
  $\Phi$ is an assignment which associates with every globally hyperbolic spacetime $(M,{\bf g})$ 
and every smooth
  contravariant tensor field on $M$, $A$ (with fixed order)
 a distribution $\Phi[{\bf g}, A]$ taking values in the algebra ${\cal W}(M,{\bf g})$. $\Phi$, is said 
 {\bf local and covariant} 
 if it satisfies the  following  \\
   {\bf Locality and Covariance requirement for differentiated fields}: {\em For any embedding $\chi$ 
   from a globally hyperbolic spacetime $(N, {\bf g}')$ 
 into another globally hyperbolic spacetime $(M, {\bf g})$ which is isometric  and causally 
preserving  it holds
\begin{eqnarray}
 {i}_\chi(\Phi[{\bf g}',A'](f)) = \Phi[{\bf g},A](f\circ \chi^{-1}) \label{lc}\:,
\end{eqnarray}
 for all $f \in {\cal D}(N)$ and all smooth vector fields $A$ on $M$, $A'$ denoting
 $(\chi^{-1})_* A\sp\rest_{\chi(N)}$. 
  The generalization to (locally and covariant) quantum field in $n$-variables and depending on 
several
 smooth contravariant vector fields is straightforward.}\\

 We conclude this part by showing that the introduced differentiated local Wick polynomial are local
 and covariant.\\

 \noindent {\bf Theorem 3.1}. {\em Take $n\in \{1,2,\ldots\}$ and, 
 for every $i\in \{1,\ldots,n\}$, take integers  $L_i= 0,1,\ldots$. Let $\Phi$ be the map which 
associates 
 with every globally hyperbolic spacetime $(M,{\bf g})$ and every class smooth contravariant 
 vector field on $M$, $\{a_{(ij)}\}_{i= 1,\ldots,n,\: j=0,\ldots L_i}$,  the (abstract)  distribution 
  $f\mapsto :\spa K_1\phi \cdots K_n\phi(f) \colon$, where $f\in {\cal D}(M)$ 
 and each $K_i$ being defined in {\em (\ref{opdif})} using the fields $a_{(ij)}$.\\ $\Phi$  is a locally 
and covariant 
  differentiated quantum field in one variable.}\\

\noindent {\em Sketch of proof.} By Def.3.2 the proof reduces to check (\ref{lc})
making use of spacetimes $(N, {\bf g}')$ and $(M, {\bf g})$ which are causal domains. 
In that case, if $H'$ and $H$ are the distributions (\ref{H}) on $N$ and $M$ respectively, 
one finds  $H'(x,y) = H(\chi(x),\chi(y))$
(provided the length scale $\lambda$ is the same in both cases). 
Representing generators $W_{\omega,n}$ in terms of generators $W_{H,n}$
as indicated in the proof of Proposition 3.1,  one also gets
that the injective $*$-algebra homomorphism $i_\chi: {\cal W}(N, {\bf g}')\to {\cal W}(M,{\bf g})$
(\ref{chi}) satisfies  $i_\chi(W_{H',n}(t)) = W_{H,n}(t\circ \chi^{-1}_n)$.
Referring to (\ref{tfdev}) and (\ref{opdif}), we adopt the notation,
$ t_n[{\bf g}, a_{(ij)},f] \stackrel {\mbox{\scriptsize  def}} {=} t_n[K_1,\ldots, K_n,f]$.
With the obtained  results  and using (\ref{phind}), (\ref{lc}) turns out to be equivalent to
$W_{H,n}( t_n[{\bf g}', (\chi^{-1})_*a_{(ij)},f] \circ \chi^{-1}_n)
= W_{H,n}( t_n[{\bf g}, a_{(ij)},f \circ \chi^{-1}])$ 
for all $n\in \bN$, $f\in {\cal D}(N)$ and all smooth  tensor fields $a_{(ij)}$ on $M$.
That identity holds because $t_n[{\bf g}, a_{(ij)},f \circ \chi^{-1}]=
t_n[{\bf g}', (\chi^{-1})_*a_{(ij)},f] \circ \chi^{-1}_n$ by the definition of distributions 
$t_n[K_1,\ldots, K_n,f]$ (\ref{tfdev})
and ${\bf g}' = \chi^* {\bf g}$.
$\Box$\\

 \noindent {\bf 3.3.} {\em The stress-energy tensor operator.}   From now on,
 $:\spa K_1\phi(x)\cdots K_n\phi(x) \colon$
  indicates the formal kernel of the one-variable distribution  $f\mapsto :\spa K_1\phi \cdots K_n\phi 
(f) \colon$.
  Using that notation and interpreting  $h\in C^\infty(M)$ as a multiplicative operator, we also  define 
\begin{eqnarray} :\spa h(x)\phi^n(x)\colon   &\stackrel {\mbox{\scriptsize  def}} {=}&
 :\spa K_1 \phi(x) \cdots K_n \phi(x)\colon \:\:\:\:\mbox{where $K_1=h$ and $K_i=I$ if $i=2,\ldots 
n$}\nonumber \\
 :\spa h(x) \nabla_X\nabla_Y  \phi^2(x)\colon &\stackrel {\mbox{\scriptsize  def}} {=}&
 2:\spa h(x)\phi(x)\:\nabla_X\nabla_Y \phi(x)\colon+   2:\spa h(x)\nabla_X\phi(x)\: 
\nabla_Y\phi(x)\colon
\nonumber\end{eqnarray}
  Let $\{Z_{(a)}\}_{a=0,1,2,3}$ be a set of {\em tetrad fields}, i.e., four smooth contravariant 
vector fields
 defined on $M$ such that ${\bf g}(Z_{(a)},Z_{(b)})(x) =
  \eta_{ab}\:,$ where $\eta_{ab}\stackrel {\mbox{\scriptsize  def}} {=}
  \eta^{ab} \stackrel {\mbox{\scriptsize  def}} {=} c_a\delta_{ab}$ everywhere
  (there is no summation with respect to a)  with $c_0=-1$  and $c_a=1$ otherwise. Making use of 
fields $Z_{(a)}$, we define
  \begin{eqnarray} :\spa h(x){\bf g}(\nabla \phi, \nabla \phi)(x)\colon &\stackrel {\mbox{\scriptsize  
def}} {=}&
   \sum_{a,b}\eta^{ab} :\spa h(x)\nabla_{Z_{(a)}} \phi(x)
 \: \nabla_{Z_{(b)}}\phi(x) \colon \nonumber \\
 :\spa h(x)\phi(x)\Delta \phi(x)\colon &\stackrel {\mbox{\scriptsize  def}} {=}&  
\sum_{a,b}\eta^{ab} :\spa h(x)\phi(x)  \:
  \nabla_{Z_{(a)}}\spa\nabla_{Z_{(b)}}\phi(x) \colon
   \nonumber           \\
  &-& \sum_{a,b}\eta^{ab} :\spa h(x)\phi(x)\: \nabla_{\left(\nabla_{Z_{(a)} }Z_{(b)}\right) } 
\phi(x)\colon\:.\nonumber 
\end{eqnarray}
   These definitions do not depend on the choice of the tetrad fields and  reduce to the usual ones
   if the field operators are replaced by  classical fields. Finally,
   $$:\spa h(x) \Delta \phi^2(x)\colon \stackrel {\mbox{\scriptsize  def}} {=}
  2:\spa h(x){\bf g}(\nabla \phi, \nabla \phi)(x)\colon + 2 :\spa h(x)\phi(x) \Delta \phi(x)\colon\:.$$
 Theorem 2.1 strongly suggests the following definition. \\

 \noindent {\bf Def. 3.3}.
 {\bf (The stress-energy tensor operator)} {\em Let $(M,{\bf g})$ be a four-dimensional smooth  
globally-hyperbolic spacetime equipped
  with a Klein-Gordon operator  {\em (\ref{operator})} with $V'\equiv 0$.
  Let  $X,Y$ be a pair of smooth vector fields on $M$.
The {\bf stress-energy tensor operator} with respect to $X,Y$ and $f\in {\cal D}(M)$, $:\spa
 T_{X,Y}(f)\colon$,
 is defined by the formal kernel
 \begin{align}
& :\spa T_{X,Y}(x)\colon \stackrel {\mbox{\scriptsize  def}} {=} \:\: :\spa 
{\nabla}_X\phi(x){\nabla}_{Y}
\phi(x)\colon - \frac{1}{2}:\spa g_{X,Y}(x){\bf g}({\nabla}\phi, {\nabla}\phi)(x)\colon
- \frac{m^2}{2}:\spa g_{X,Y}(x)\phi^2(x)\colon \nonumber \\
& - \frac{1}{2}:\spa g_{X,Y}(x)R(x)\phi^2(x)\colon\sp_\lambda
 +\xi:\spa \left( R_{X,Y}(x) -\frac{1}{2}g_{X,Y}(x) R(x)\right) \phi^2(x)\colon + \xi
  :\spa g_{X,Y}(x)
 \Delta {\phi^2(x)}\colon \nonumber\\
&  - \xi :\spa g_{X,Y}(x) \nabla_X\nabla_{Y}{\phi^2(x)}\colon
 + \frac{1}{3}:\spa g_{X,Y}(x) \phi(x) P\phi(x)\colon
 \:,    \label{tensoroperator}
 \end{align}
 where     $g_{X,Y}(x)  \stackrel {\mbox{\scriptsize  def}} {=} {\bf g}(X,Y)(x)$,    
 $R_{X,Y}(x)  \stackrel {\mbox{\scriptsize  def}} {=} {\bf R}(X,Y)(x)$, ${\bf R}$
 being the Ricci tensor
 and
 \begin{eqnarray}
 :\spa g_{X,Y}(x) \phi(x) P\phi(x)\colon    \stackrel {\mbox{\scriptsize  def}} {=}
   - :\spa g_{X,Y}(x)  \phi(x) \Delta \phi(x) \colon +
 :\spa g_{X,Y}(x) (R(x)+ m^2)\phi^2(x)\colon        \label{phiPphi}\:.
 \end{eqnarray}}

 \noindent {\em Remarks}. {\bf (1)} We have introduced the, classically vanishing,  
term $:\spa g_{X,Y}(x)  \phi(x) P \phi(x) \colon$. Its presence  is crucial
 to obtain the conservation of the stress-energy tensor operator using the analogous property of the 
point-splitting 
renormalized stress-energy tensor as done in the proof of the theorem below. \\
{\bf (2)} The given definition depends on the choice of a length scale $\lambda$ present in
the distribution $H$ used to define the local Wick products of fields operators.\\

 To conclude our analysis we analyze the interplay between the above-introduced
 stress-energy tensor operator and the point-splitting procedure discussed in the Section 2.
 Concerning the issue of the conservation of the stress-energy tensor, we notice in advance that,
 if $T$ is a second-order covariant symmetric tensor field,
 \begin{eqnarray}
 - \int_M \sp f\: \left( \nabla\cdot T \right)_X d\mu_{\bf g}
  =  \int_M \sp f \:T_{\nabla \otimes X} d\mu_{\bf g}
  + \sum_{a,b} \eta^{ab} \int_M\left\{  T_{Z_{(a)},X} \nabla\cdot \left(fZ_{(b)}\right) +
f\: T_{X, \nabla_{Z_{(a)}} Z_{(b)}}\right\} d\mu_{\bf g} \label{div1}
 \end{eqnarray}
for all $f\in {\cal D}(M)$ and  all smooth contravariant vector fields $X$ on $M$.
Above $(\nabla \cdot T)_X = (\nabla^\mu T_{\mu\nu}) X^\nu$ and $T_{\nabla \otimes X} =
 T_{\mu\nu} \nabla^\mu
X^\nu$ in the abstract index notation. Therefore the conservation requirement $\nabla \cdot T
\equiv 0$ is equivalent to the requirement that the right-hand side of (\ref{div1})
vanishes for all $f\in {\cal D}(M)$ and  smooth contravariant vector fields $X$ on $M$.

  We have a following conclusive
 theorem where, if $\nu$ is a quasifree Hadamard state,
$:\spa{T}_{\nu\:X,Y}(f)\colon$,
$:\spa\hat{\phi}_\nu^2(f)\colon$ respectively
represent $:\spa{T}_{X,Y}(f)\colon$ and  $:\spa{\phi}^2(f)\colon$ in ${\cal W}_{\nu}(M,{\bf g})$ 
in the sense of (b) in Proposition 3.2.\\

  \noindent {\bf Theorem 3.2.} {\em Let $(M,{\bf g})$ be a four-dimensional smooth globally-
hyperbolic spacetime equipped
  with a Klein-Gordon operator {\em (\ref{operator})} with $V'\equiv 0$. Let $\lambda >0$ be
the scale length used to define local Wick products of fields operators and $\{Z_{(a)}\}_{a=1,\ldots,4}$ 
a set of tetrad fields.
 Considering the given definitions, the statements below hold for every  $f \in {\cal D}(M)$.\\
  {\bf (a)}  For every $h\in C^\infty(M)$, $:\spa h(x) \phi(x) P \phi(x) \colon$ does not depend on 
$\lambda$ and  turns out to be
   a smooth function. In particular if  $U_2(x,x)$ is  defined as in the Appendix A,
  \begin{eqnarray}
  :\spa h \phi P \phi(f) \colon =  \frac{3}{2\pi^2} \left( \int_M h(x) U_{2}(x,x) f(x) d\mu_{\bf g}(x) 
\right) 1    \label{universal}
  \end{eqnarray}
  {\bf (b)} Take a quasifree Hadamard state $\nu$ and let $\omega$ be any (not necessarily quasifree) 
Hadamard 
  state represented by $\Psi_\omega\in D_{\nu}\subset {\cal H}_{\nu}$ in a GNS representation  
  of $\nu$. For every pair of contravariant vector fields
 $X,Y$, it holds
  \begin{align}
    \left\langle\Psi_{\omega}, :\spa\hat{T}_{\nu XY}(f)\colon 
\Psi_{\omega} \right\rangle_{\nu}  =&
    \int_M \langle\hat{T}_{XY}(z)\rangle_{\omega} f(z) d\mu_{\bf g}(z) 
     \label{equivalence}   \:,   \\
     \left\langle\Psi_{\omega},  :\spa\hat{\phi}_{\nu}^2(f)\colon \Psi_{\omega} \right\rangle_{\nu}
 =&
    \int_M \langle\hat{\phi}^{2}(z)\rangle_{\omega} f(z) d\mu_{\bf g}(z) 
     \label{equivalence2}\:,
  \end{align}
   
$\langle\hat{T}_{XY}(z)\rangle_{\omega}=\langle\hat{T}_{\mu\nu}(z)\rangle_{\omega}X^\mu(x) 
Y^{\nu}(z)$
   and $\langle\hat{\phi}^{2}(z)\rangle_{\omega}$ denoting the fields obtained by the 
   point-splitting procedure {\bf Def.2.1}.\\
  {\bf (c)} The stress-energy tensor operator is {\bf conserved}, i.e.,  for every contravariant vector 
field  $X$  on $M$
  it holds
  \begin{eqnarray}
    :\spa \left( \nabla\cdot T \right)_X(f) \colon  = 0\:,
  \label{conservation2}
  \end{eqnarray}
  where, following {\em (\ref{div1})},
  \begin{eqnarray}:\spa \left( \nabla\cdot T \right)_X(f) \colon \stackrel {\mbox{\scriptsize  def}} {=} 
-:\spa T_{\nabla \otimes X}(f) \colon 
  - \sum_{a,b} \eta^{ab} \left\{  :\spa T_{Z_{(a)},X} \left(\nabla\cdot \left(fZ_{(b)}\right)\right 
)\colon +
:\spa T_{X, \nabla_{Z_{(a)}} Z_{(b)}}(f) \colon\right\}\:. \label{div2}\end{eqnarray}
  {\bf (d)} The trace of the stress-energy tensor operator satisfies
   \begin{eqnarray}
   \sum_{a,b} \eta^{ab}  :\spa T_{Z_{(a)},Z_{(b)}}(f)\colon  &=& \frac{6\xi-1}{2} :\spa \Delta 
\phi^2(f)\colon -
   :\spa (m^2 + \xi R) \phi^2(f)\colon \nonumber\\
   &+& \frac{1}{2\pi^2} \int_M U_2(x,x) f(x) d\mu_{\bf g}(x) \:\: 1
              \label{trace2'} \:.\end{eqnarray}
  {\bf (e)} If $0< \lambda' \neq \lambda$, with obvious notation,
   \begin{eqnarray}
:\spa T_{X,Y}(f)\colon\sp_{(\lambda)}\:\: - \::\spa T_{X,Y}(f)\colon\sp_{(\lambda')} =
\ln\left(\frac{\lambda'}{\lambda}\right) \int_M t_{X,Y}(x) f(x) d\mu_{\bf g}(x)\:\: 
1\:,\label{difference2}
\end{eqnarray}
where the smooth, symmetric, conserved  tensor field $t$ is that introduced in  {\em 
(\ref{difference})}.}\\

 \noindent {\em Proof.} We start by proving (\ref{equivalence2}) which is the simplest item. 
 It is obvious by Def.3.2
that we may reduce to consider $f\in {\cal D}(N)$ where $N\subset M$ is a causal domain. We have 
$$\left\langle\Psi_{\omega}, :\spa\hat\phi_{\nu}^2(f)\colon\Psi_{\omega}\right\rangle_{\nu} =
 \lim_{j\to \infty}
 \left\langle\Psi_{\omega}, :\spa\hat\phi^2_{\nu}(s_j)\colon \Psi_{\omega}\right\rangle_{\nu}\:,$$
where $\{s_j\} \subset {\cal D}(N^2)$ is a sequence of smooth functions which converge to 
$t_2(I,I,f) = f\delta_2$
in the H\"ormander pseudo topology in a closed conic set in $N\times (\bR^4\setminus\{0\})$
 containing $WF(f\delta_2)$. Such a sequence does exist
by Theorem 8.2.3 of \cite{hor1}. Above we have used the continuity of the scalar product as well as
the continuity of the map $t_2(I,I,f) \mapsto :\spa\hat\phi^2_{\nu}(f)\colon\Psi_{\omega}$ since 
$\Psi_{\omega}\in D_{\nu}$.
On the other hand we may choose each $s_j$ of the form $\sum_j c_j h_j\otimes h_j'$,
where the sum is finite, $c_j\in \bC$ and $h_j,h'_j\in {\cal D}(N)$. This is because, using 
Weierstrass' theorem 
on uniform
approximation by means of polynomials in $\bR^m$, it turns out that  the space of finite linear 
combinations
$h\otimes h'$ as above is dense in ${\cal D}(N\times N)$ in its proper seminorm-induced topology
(viewing $N$ as a subset of $\bR^4$ because of the presence of global coordinates). We leave the 
trivial details 
to the reader.
With that choice
one straightforwardly finds
$$ \left\langle\Psi_{\omega}, :\spa\hat\phi^2_{\nu}(s_j)\colon
 \Psi_{\omega}\right\rangle_{\nu} = (G^{(1)}_{\omega}-H)(s_j)=
s_j(G^{(1)}_{\omega}-H)\:,$$
where we have used the fact that both $G^{(1)}_{\omega_)}-H$ and $s_j$ are  smooth. Since the 
convergence
in the H\"ormander pseudotopology imply the convergence in ${\cal D}'(N)$, we finally get
$$\left\langle\Psi_{\omega}, :\spa\hat\phi^2_{\nu}(f)\colon
 \Psi_{\omega}\right\rangle_{\nu} = \lim_{j\to \infty}s_j(G^{(1)}_{\omega}-H)=
\int_{N\times N } (G^{(1)}_{\omega}-H) (x,y) f(x) \delta_2(x,y) d\mu_{\bf g}(x) d\mu_{\bf 
g}(y)$$
The achieved result can be re-written in a final form  taking (\ref{phi2}) into account and noticing that 
$Z_{n} -H$
is $C^n(N\times N)$, it vanishes  with all of the derivatives up to the order $n$ for $x=y$ and $n$ 
may be fixed arbitrarily
large. By this way we get
$$ \left\langle\Psi_{\omega},  :\spa\hat{\phi}^2_{\nu}(f)\colon \Psi_{\omega} \right\rangle_{\nu} =
    \int_M \langle\hat{\phi}^{2}(x)\rangle_{\omega} f(x) d\mu_{\bf g}(x)$$
 which is nothing but our thesis. Using the same approach one may prove (\ref{equivalence}) as well 
as
$$ \left\langle\Psi_{\omega},  :\spa h\hat\phi_{\nu} P\hat{\phi}_{\nu}(f)\colon \Psi_{\omega} 
\right\rangle_{\nu} =
    \int_M h(x) \langle\hat\phi (x)P\hat\phi(x)\rangle_{\omega} f(x) d\mu_{\bf g}(x)\:,$$
where $f\in {\cal D}(N)$ and $h\in C^\infty(M)$.
In other words, by Lemma 2.1,
$$ \left\langle\Psi_{\omega},  :\spa h\hat \phi_{\nu}P\hat{\phi}_{\nu}(f)\colon \Psi_{\omega} 
\right\rangle_{\nu} =
    -\int_M  h(x) c_4 U_2(x,x) f(x) d\mu_{\bf g}(x)\:.$$
The right hand side does not depend on $\Psi_{\omega}$ which, it being Hadamard as $\nu$ (but not 
necessarily 
quasifree), may range in the dense  subspace of the Fock space ${\cal H}_\omega$ containing 
$n$-particle states with smooth modes \cite{hw}.
Finally, using the fact that the Hilbert space is complex one trivially gets the operator identity on 
${D}_\omega$
 $$ :\spa h\hat \phi_{\nu} P\hat{\phi}_{\nu}(f)\colon =  -\int_M  h(x) c_4 U_2(x,x) f(x) 
 d\mu_{\bf g}(x) I\:.$$
By Def.3.2. such an identity holds true also for $f\in {\cal D}(M)$, then
$$ :\spa h\phi P{\phi}(f)\colon = \alpha^{-1}_{\nu}\left(:\spa h\hat \phi_{\nu} 
P\hat{\phi}_{\nu}(f)\colon\right)
= \int_M  h(x) c_4 U_2(x,x) f(x) d\mu_{\bf g}(x)\:,$$
because $\alpha$ is an algebra isomorphism. We have proven the item (a). The items
(e) and (d) may be proven analogously starting from (\ref{equivalence}),
in particular (d) is  a direct consequence of (b) in Theorem 2.1.
Let us  prove the conservation of the stress-energy tensor operator (\ref{conservation2}).
To this end, we notice that (\ref{equivalence}) together with the item (a) of Theorem 2.1 for 
$V'\equiv 0$
by means of the procedure used to prove (\ref{universal}) implies the operator identity on ${\cal 
H}_{\nu}$
\begin{eqnarray}
  \sp :\spa \hat T_{\nu\: \nabla\otimes X}(f) \colon
+\sum_{a,b} \eta^{ab} \left\{:\spa \hat T_{\nu\: Z_{(a)},X}\left(\nabla \cdot (f Z_{(b)})\right)\colon 
+
  :\spa \hat T_{\nu\:X, \nabla_{Z_{(a)}} \spa Z_{(b)}}(f)\colon\right\}  = 0      
  \label{defdiv2}\:,
  \end{eqnarray}
 for any
Hadamard state $\omega$, $X,Y$ smooth vector fields and $Z_{(a)}$ tetrad fields.
This identity entails (\ref{conservation2}) by applying  $\alpha^{-1}_{\nu}$ on both sides.
$\Box$\\

\section{Summary and final comments.} We have shown that a definition of stress-energy tensor 
operator
in curved spacetime is possible in terms of local Wick products of field operators only by adding 
suitable
terms to the classical form of the stress energy tensor. Such a
definitions seems to be quite reasonable and produces results in agreement with well-known 
regularization procedures of
averaged quantum observables. The added terms $:\spa h \phi P \phi (f)\colon$ in the form of the 
stress-energy
tensor operator enjoy three remarkable properties. (1) They classically vanish,
(2) they are written as local Wick products of field operators,
(3) they are in a certain sense {\em universal}, i.e., they do not depend on the scale $\lambda$ and 
belong to the commutant 
of the algebra
${\cal W}(M, ${\bf g}$)$.\\
The issue whether or not it could be  possible to define a stress-energy tensor operator free from these 
terms is related to
the issue of the existence of Hadamard singular  bidistributions, defined locally and somehow 
"determined by the local geometry only".
The positiveness seems not to be a requirement strictly necessary at this level.
The appearance of the so-called conformal anomaly shared by the various regularization
techniques and related to the presence of the found  terms could be in contrast to the existence of
such local bisolutions. However, no proof, in any sense, exists  in literature to the knowledge of the 
author.\\
 As a final comment we suggest that the  universal terms  $:\spa h \phi P \phi(f) \colon$ or similar 
terms may be useful
 in studying other conservation laws within the approach  based on local Wick products, e.g., 
conserved currents in
 (non-)Abelian gauge theories and related anomalies.\\

 \noindent {\bf Acknowledgments}.\\ I am grateful to Romeo Brunetti and Klaus Fredehagen  
 for helpful discussions and comments.

 \section*{Appendix A.}
 If $(M,{\bf g})$ is a smooth Riemannian or Lorentzian manifold, an open set $C\subset M$
 is said a {\em normal convex} neighborhood
 if
 there is a open set $W \subset TM$, $W=\{ (q,v)\:|\: q\in C, v\in S_q\}$,  $S_q\subset T_qM$ being
 a starshaped open neighborhood of the origin, such that $exp\spa\rest_W : (q,v) \mapsto
 exp_q v$ is a diffeomorphism onto $C\times C$. It is clear that $C$ is connected and  there is only 
one geodesic segment
 joining any pair $q,q'\in C$, completely contained in $C$, i.e., $t\mapsto exp_q(t ((exp_q)^{-
1}q'))$
 $t\in [0,1]$. Moreover if $q\in C$, $\{e_\alpha|_q\}\subset T_qM$ is a basis,
  $t = t^\alpha e_\alpha|_q \mapsto exp_q(t^\alpha e_\alpha|_q)$, $t\in S_q$ defines a set of 
 coordinates on $C$ centered in $q$ which is called {\em normal Riemannian coordinate 
system}  centered in   $q$.
 In $(M,{\bf g)}$ as above, $s(x,y)$ indicates the squared geodesic distance of $x$ from $y$:
 $s(x,y) \stackrel {\mbox{\scriptsize  def}} {=}{\bf g}_x(exp_x^{-1}y,exp_x^{-1}y)$.  By 
definition
 $s(x,y)=s(y,x)$ and $s$ turns out to be  smoothly defined on $C\times C$ if $C$ is a convex normal 
neighborhood.
 The class of the convex normal neighborhood of a point $p\in M$ defines a fundamental system of
 neighborhoods of $p$ \cite{friedlander,BEE}.
 With the signature $(-,+,\cdots,+)$, we have  $s(x,y) > 0$ if the points are space-like separated,
 $s(x,y)<0$ if the points are time-like related and $s(x,y)=0$ if the points are light related.
In Euclidean manifolds $s$ defined as above  is everywhere nonnegative.
 
 The distribution $Z^{(+)}_{n}$ is defined by the following  integral kernel
in the sense of the usual $\epsilon\to 0^+$ prescription.
 \begin{align}
 \beta^{(1)}_D \: \frac{{U}(x,y) }{{s_{\epsilon,T}}^{D/2-1}(x,y)}+
 \beta^{(2)}_D {V}^{(n)}(x,y) \: \ln
 \frac{s_{\epsilon,T}(x,y)}{\lambda^2}&\:\:\:\:\:\:   \mbox{if $D$ is even},   \label{W2'}    \\
\beta^{(1)}_D
 \frac{{T}^{(n)}(x,y) }{s_{\epsilon,T}^{D/2-1}(x,y)} &\:\:\:\:\:\:   \mbox{if $D$ is odd}  \:.   
\label{W2''}
\end{align}
$s^k(x,y) \stackrel {\mbox{\scriptsize  def}} {=}(s(x,y))^k$.
The cut branch in the logarithm is fixed along the negative real axis, moreover
  $s_{\epsilon, T}  \stackrel{\mbox{\scriptsize  def}} {=} s(x,y) + i\epsilon (T(x)-T(y)) 
+\epsilon^2$, $T$
being any global temporal function defined on $M$ increasing  toward the future.
  The distributio $Z^{(+)}_{n}$
 does not depend on the choice of $T$   (see \cite{kw} and Appendix A3 of \cite{sv}). 
 $\lambda \in \bR$ is a length scale arbitrarily fixed.
 \begin{align}
 \beta^{(1)}_D &\stackrel {\mbox{\scriptsize  def}} {=} 
(-1)^{\frac{D+1}{2}} \frac{\pi^{\frac{2-
D}{2}}}{ 2\Gamma(\frac{4-D}{2})} \:\:\:\:\:\:\:\:\mbox{for $D$ odd,}
  \:\:\:\:\:\:\:\:\:\:\:\:\:\:\:\:
 \beta^{(1)}_D \stackrel {\mbox{\scriptsize  def}} {=} -\frac{\Gamma(\frac{D}{2}-1)}{2 
\pi^{\frac{D}{2}}} \:\:\:\:\:\:\:\:\mbox{for $D$ even,}\\
 \beta^{(2)} &\stackrel {\mbox{\scriptsize  def}} {=}    (-1)^{\frac{D}{2}} \frac{2^{1-D}  }{\pi^{\frac{D}{2}}\Gamma(\frac{D}{2})}\:.
 \end{align}
 $U,V^{(n)},T^{(n)}$ admit the following expansions in 
powers of $s(x,y)$. If $D$ is even
\begin{align}
 U(x,y) &\stackrel {\mbox{\scriptsize  def}} {=} \Theta_D \sum_{k=0}^{(D-4)/2} \frac{1}{(4-
D|k)} \:U_{k}(x,y)\: s^k(x,y) \:,
\label{U} \\
  V^{(n)}(x,y) &\stackrel {\mbox{\scriptsize  def}} {=}  \left(2\left|\frac{D}{2}-1\right)\right. 
\:\sum_{k=0}^n \frac{1}{2^k
k!} U_{\frac{D}{2}+k-1}(x,y)s^k(x,y) \:, \label{V}
 \end{align}
 where $\Theta_2=0$ and $\Theta_D =1$ if $D>2$, and
 \begin{align}
   T^{(n)}(x,y) &\stackrel {\mbox{\scriptsize  def}} {=} \sum_{k=0}^{n+(D-3)/2}   \frac{1}{(4-
D|k)}\:
U_{k}(x,y)\:s^k(x,y)      \:, \label{T}
\end{align}
if $D\geq 3$ is odd.  
$(\alpha|0) \stackrel {\mbox{\scriptsize  def}} {=}1$ and $(\alpha | k) \stackrel {\mbox{\scriptsize  
def}} {=} \alpha
(\alpha+2)\cdots (\alpha +2k -2)$. For any  open convex  normal neighborhood $C$ in
 $M$ there is exactly one sequence   of $C^{\infty}(C\times C)$ real valued functions
 $U_{k}$, used in the expansions above, satisfying the differential equations on ${C}\times {C}$:
\begin{eqnarray}
P_xU_{k-1}(x,y) +{\bf g}_x({\nabla}_{(x)} s(x,y), U_{k}(x,y)) + (M(y,x) +2k) U_{k}(x,y) = 0 
\label{recurrence}\:,
\end{eqnarray}
 with the initial conditions
$U_{-1}(x,y) = 0$ and $U_{0}(x,x) = 1$.
 The function $M$ is defined as
$M(x,y) \stackrel {\mbox{\scriptsize  def}} {=} \frac{1}{2}\Delta_x s(x,y) - D\:,$
with $D$ is the dimension of the manifold.  The proof of existence and uniqueness is trivial 
using normal coordinates centered in $x$.
The coefficients $U_{k}(x,y)$ can be defined, by the same way, also if the metric is Euclidean.
They coincide, barring numerical factors, with the so called {\em  Hadamard-Minakshisundaram-
DeWitt-Seeley coefficients}.
If $C',C$ are convex normal
neighborhoods and $C'\subset C$, the restriction to $C'$ of each  $U_{k}$ defined in $C$
 coincides with the corresponding  coefficient directly defined on $C'$.
There is a wide literature on coefficients $U_k$,
 in relation with heat-kernel theory and $\zeta$-function regularization technique \cite{mzp}.

As in {\bf 2.2}  $Z(x,y)$ indicates the kernel of $Re (Z^{(+)})$ which is smooth 
for $s(x,y)\neq 0$ in every convex normal neighborhood $C_z\ni x,y$. 
It is possible to show that the coefficients are symmetric, i.e., if $x,y\in C$,
 $U_{n}(x,y)= U_{n}(y,x)$ \cite{msl} and thus, since $s(x,y)=s(y,x)$, it also holds
 $$Z_{n}(x,y) = Z_{n}(y,x)$$ for any $n\geq 0$ and $s(x,y)\neq 0$.

The recurrence relations (\ref{recurrence}) have been obtained by requiring that
the sequence $Z_{n}$ defines a local $y$-parametrized
 "approximated solution" of  $P_xS(x,y) = 0$ \cite{garabedian}.
That solution is exact if one takes the limit $Z = \lim_{n \to \infty} Z_{n}$
of the sequence provided the limit exist.
This happens in the analytic case, but in the smooth general case the sequence may diverge.
Actually, in order to produce an approximated/exact solution
for  $D$ is even, a smooth part $W$ has to be added to
$Z$, $S = Z+W$, and also $W$ can be expanded in powers of $s$ \cite{garabedian}.  Differently 
from
the expansion of $Z$ which is completely determined by the geometry and the operator $P$, the 
expansion of
$W$ depends on its first term  $W_0$ (corresponding to $s^0$) and there is no natural choice of
$W_0$ suggested by $P$ and the local geometry.   Finally if $D$ is even and for whatever choice of
$W_0$ there is no guarantee for producing a function $Z+W$ (provided the limits of corresponding 
sequences
exist) which is solution of
field equations  in both arguments: This is because in general $W(x,y)\neq W(y,x)$ also if 
$W_0(x,y)$ is symmetric
\cite{wald78}.  \\

 \section*{Appendix B.}

 Referring to {\bf 2.2}, the properties
 (b) and (c) respectively imply the relations in $M\times M$
 \begin{eqnarray}
G_{\omega}^{(1)}(x,y) &=& G^{(1)}_\omega(y,x) \label{symmetry}\:, \\
P_xG_{\omega}^{(1)}(x,y) &=& P_yG_{\omega}^{(1)}(x,y) = 0 \label{bisolution} \:,
\end{eqnarray}
which hold when $x\neq y$  are not light-like related.
These relations  are useful in the following.

 {\em Proof} of {\bf Lemma 2.1}.
 (a) In the following we take advantage of the identity, where $f$ and $g$ are  $C^2$ functions,
 $P(f g) =f Pg  - (\Delta f)g - 2 {\bf g}({\nabla}f, {\nabla }g)\:.$
 Suppose $D > 2$ even, using the identity above and the definition of $Z_{n}$,
 one finds,  for either $x,y$ time-like related or space-like separated,
 \begin{align}
 P_xZ_{n}(x,y) & = \beta^{(1)}_D\left(\frac{P_xU(x,y)}{s^{D/2-1}} -  U(x,y) \Delta_{(x)} s^{1-
D/2} -
 2{\bf g}_x( {\nabla}_{(x)}s^{1-D/2}, \nabla_{(x)} U(x,y)) \right)\nonumber\\
 & +    \beta^{(2))}_D \left[ (P_x V^{(n)}(x,y)) \ln \frac{|s|}{\lambda^2} - V^{(n)}(x,y) 
\Delta_{(x)} \ln \frac{|s|}{\lambda^2}
  -2 \frac{{\bf g}_x(\nabla_{(x)} s, \nabla_{(x)}  V^{(n)}(x,y))}{s} \right]  \nonumber \:.
 \end{align}
 Using (\ref{recurrence}) for $n\geq 1$ we have
  \begin{align}
 P_xZ_{n}(x,y)  =  - \beta^{(2)}_D
 \left[  -(\ln \frac{|s|}{\lambda^2})P_x V^{(n)}(x,y)+ (\Delta_{(x)}  \ln \frac{|s|}{\lambda^2} )  
{V'}^{(n)}(x,y)+
 2\frac{{\bf g}_x(\nabla_{(x)}s, \nabla_{(x)} {V'}^{(n)}(x,y))}{s} \right]\:,  \nonumber 
 \end{align}
 where
$${V'}^{(n)}(x,y) \stackrel {\mbox{\scriptsize  def}} {=}  \sum_{k=1}^n V_k(x,y)  s^k(x,y)$$
with $$V_k(x,y) \stackrel {\mbox{\scriptsize  def}} {=}  \left(2\left|\frac{D}{2}-1\right)\right. \: 
\frac{1}{2^k
k!}\:U_{\frac{D}{2}+k-1}(x,y)\:.$$ Expanding the derivatives and using (\ref{recurrence}) once 
again,  if $n\geq 1$, one
gets that ,
 $-(\beta^{(2)}_D)^{-1}P_xZ_{n}(x,y)$ equals
 \begin{eqnarray}
  & &s  (\frac{\Delta_{(x)} s(x,y)}{s} -\frac{4}{s}) V_1(x,y) + 2
 \frac{ {\bf g}_x(\nabla_{(x)} s(x,y), \nabla_{(x)} s(x,y))}{s}V_1(x,y)+   2 {\bf 
g}_x({\nabla}_{(x)}s, \nabla_{(x)} V_1(x,y))\nonumber \\
&+&  |s|^{n}O_{1,n}(x,y)\ln\frac{|s|}{\lambda^2}  +  |s|^{n-1}O_{2,n}(x,y)+ |s|^{n-
1/2}O_{3,n}(x,y)\label{limit} \:,
 \end{eqnarray}
where  $O_{k, n}$ are smooth in a neighborhood of  $(z,z)$ and the last two terms appear
for $n>1$ only. Using
 $ {\bf g}_x(\nabla_{(x)} s(x,y), \nabla_{(x)} s(x,y)) = 4s(x,y)$,
one finds
 \begin{eqnarray}
  -(\beta^{(2)}_D)^{-1} P_xZ_{n}(x,y) &=& (\Delta_{(x)} s(x,y) -4) V_1(x,y) +
 8V_1(x,y)+   2 {\bf g}_x({\nabla}_{(x)}s, \nabla_{(x)} V_1(x,y))\nonumber \\
&+&  |s|^{n}O_{1,n}(x,y)\ln\frac{|s|}{\lambda^2}  +  |s|^{n-1}O_{2,n}(x,y)+ |s|^{n-
1/2}O_{3,n}(x,y)\:,\label{even}
 \end{eqnarray}
and thus, since    $\Delta_{(x)} s(x,y) \to 2D$ and $\nabla_{(x)}s(x,y) \to 0$ as $(x,y)\to (z,z)$,
 \begin{eqnarray}
  \lim_{(x,y)\to (z,z)} P_x  Z_{n}(x,y) =   c_D
   \:  {U}_{D/2}(z,z)\:,   \nonumber
 \end{eqnarray}
 which is a part of (\ref{lemma}) for $D$ even. $Z_{n}(x,y)= Z_{n}(y,x)$ implies 
the remaining identity in (\ref{lemma}).
   If $D=2$, $x,y$ are either time-like  related or space-like separated and $n\geq 1$, the proof is 
essentially the same. 
One directly finds
 \begin{align}
 P_xZ_{n}(x,y)  =  - \beta^{(2)}_2
 \left[ (-P_x V^{(n)}(x,y)) \ln \frac{|s|}{\lambda^2}+ V^{(n)}(x,y) \Delta_{(x)}  \ln 
\frac{|s|}{\lambda^2}  +
 2\frac{{\bf g}_x({\nabla}_{(x)}s, \nabla_x V^{(n)}(x,y))}{s} \right] \nonumber\:,
 \end{align}
 with   $$V^{(n)}(x,y) \stackrel {\mbox{\scriptsize  def}} {=}  \sum_{k=0}^n V_k(x,y)  
s^k(x,y)$$
and
 $$V_k(x,y) \stackrel {\mbox{\scriptsize  def}} {=} \left(2\left|\frac{D}{2}-1\right)\right. \: 
\frac{1}{2^k k!}
\:U_{\frac{D}{2}+k-1}(x,y)\:.$$
 Using  $V_0 = U_0$ ($D=2$) and     (\ref{recurrence}) for $k=0$,  one gets (\ref{even}) once 
again.\\
    If $D$ is odd, $n\geq 1$ and  $x,y$ are either  space-like separated or time-like related,  
(\ref{recurrence})  
entails
 $$P_xZ_{n}(x,y)  = \theta(s(x,y)) |s(x,y)|^{n-1/2}  O_n(x,y)  \nonumber \:.$$
    where  $O_n$ is smooth in a neighborhood of  $(z,z)$.  Therefore,
 \begin{eqnarray}
  \lim_{(x,y)\to (z,z)} P_x  Z_{n}(x,y) =    0  \:.  \label{odd}
 \end{eqnarray}
 which is a part of  (\ref{lemma}) for $D$ odd the other part is a trivial consequence
 of the symmetry as above. Notice that the proof shows also that the limit is uniform
 in the three treated  cases because $|s(x,y)|$ uniformly  tends to $0$ as $(x,y)\to (z,z)$. \\
 (b) The proof follows  a very similar procedure
 as that used in the proof of (a).  One obtains  that 
$\lim_{(x,y)\to (z,z)} P_x  \nabla^\mu_{(y)}  Z_{n}(x,y)$
 equals
  \begin{eqnarray}
 - \beta_D^{(2)}
\left[(2D- 4) \nabla^\mu_{(y)} V_{1}(x,y) +
 8 \:{\nabla}^\mu_{(y)} V_{1}(x,y) -
 4 \:{\nabla}^\mu_{(x)} V_{1}(x,y)\right]_{x=y=z}\:. \label{finis}
\end{eqnarray}
(One has to differentiate  (\ref{even})   with
 respect to $\nabla_{(y)}^\mu$ and   use $\nabla_{(y)}^\mu \Delta_x s(x,y)) |_{x=y=z} = 0$ and
  $\nabla_{(x)\mu}\nabla_{(y)\nu} s(x,y)|_{x=y=z} = -2 g_{\mu\nu}(z)$.)
  Finally one notices that $V_1$ is proportional to $U_{D/2}$ and,
   since $U_{n}(x,y) = U_{n}(y,x)$, it also holds
   $\nabla^{\alpha}_{(z)}V_{1}(z,z)= 2\nabla^{\alpha}_{(y)}V_{1}(x,y)|_{x=y=z}\:.$
 Using that in (\ref{finis}) 
    the thesis (b) arises.         For $D=2$ and $D$ odd the proof is the same with trivial modifications. 
  (c) The proof directly follows from  (a) and (\ref{bisolution}).
 $\Box$  \\

 \noindent {\em Proof} of {\bf Theorem 2.1}.  (a) $(x,y) \mapsto G_\omega^{(1)}(x,y)-
Z_{n}(x,y)$
 is $C^n$ in a whole neighborhood of $z$ (also for light-like related arguments). Since we want to apply the second order operator
 ${\cal D}^{(\eta)}_{(z)\mu\nu}$ to it, we need to fix $n\geq 2$, however we also want to derive
 the obtained stress-energy tensor and thus we need $n\geq 3$. With $n\geq 3$  the map above
 is $C^n$ at least and $z \mapsto \langle \hat{T}_{\mu\nu}^{(\eta)}(z) \rangle_{\omega}$
 is $C^{n-1}$.   Finally, since the latter map do not depend on $n$ it must be $C^\infty$
 also if $n \geq 3$ is finite.
 The independence from $n$ is a consequence of
  $\lim_{(x,y)\to (z,z)}  \Delta_{n,n'}(x,y) =0$ where
  $ \Delta_{n,n'}(x,y) \stackrel {\mbox{\scriptsize  def}} {=} Z_{n}(x,y)-Z_{n'}(x,y)$
  which holds true for any pair $n,n' \geq 3$   as the reader may straightforwardly check
  and prove by induction.
  The remaining part of (a) may be proven as follows. For any $C^3$ function $(x,y) \mapsto 
\Gamma(x,y)$
   symmetric under interchange of $x$ and $y$ we have the identity
   \begin{eqnarray}
   \nabla^{\mu}_{(z)}\left({\cal D}^{(\eta)}_{(z)\mu\nu} \Gamma(x,y)|_{x=y=z} \right) &=& - 
P_x\nabla_{(y)\nu} \Gamma(x,y)|_{x=y=z}
   + \eta  \nabla_{(z)\nu} \left( P_x \Gamma(x,y)|_{x=y=z}\right) \nonumber \\
    &-& \frac{1}{2}\Gamma(z,z) \nabla_\nu V'(x)
  \label{identity}\:.
   \end{eqnarray}
    Indeed, if $\phi$ does not satisfy the field equation,   (\ref{conservationT}) reads,
 \begin{eqnarray}
 \nabla^\alpha T_{\alpha\beta}(x) = - {(P \phi)(x)} \nabla_{\beta}\phi(x)
 -  \frac{1}{2}\phi^2(x) \nabla_{\beta} V'(x)  \label{conservationT'}\:,
 \end{eqnarray}
 Such an identity can be obtained by using the form of the stress-energy tensor and the
 symmetry of $\Gamma(x,y) = \phi(x)\phi(y)$ only.    So it holds
 true for each symmetric sufficiently smooth map $(x,y) \mapsto \Gamma(x,y)$.
   The proof of (\ref{identity}) is nothing but the proof of (\ref{conservationT'}) taking the added
  term proportional to $\eta$ into account.     Then put $\Gamma(x,y) = G_\omega^{(1)}(x,y)-
Z_{n}(x,y)$   into
(\ref{identity})
  with $n\geq 3$, this is allowed because $Z_{n}$ is symmetric by construction (see 
Appendix A) and
  $G_\omega^{(1)}$ satisfies (\ref{symmetry}). The first line of right-hand side of (\ref{identity}) reduces to
  $$ \lim_{(x,y)\to (z,z)} P_x\nabla_{(y)\nu} Z_{n}(x,y)
   - \eta \nabla_{(z)\nu} \lim_{(x,y)\to (z,z)} P_x Z_{n}(x,y)$$
   because of  (\ref{bisolution}).   Both terms above can be computed by Lemma 2.1
   finding
   $$\nabla^\mu \langle \hat{T}^{(\eta)}_{\mu\nu}(z) \rangle_{\omega} = \delta_D (k_D - \eta c_D) 
\nabla_\nu U_{D/2}(z,z)
   - \frac{1}{2} \langle \hat{\phi}^2(z) \rangle_{\omega} \nabla_\nu V'(z)\:.$$
   The former term in the right hand side vanishes if and only if  $\eta= k_D/c_D$, i.e., $\eta= 
\eta_D$.
   This concludes the proof of (a).
   (b) Directly from the form of the stress-energy tensor, one finds that, if $P\phi \neq 0$, 
(\ref{trace}) reads
    \begin{eqnarray}
g_{\alpha\beta}(x){T}^{\alpha\beta}(x) &=& \left[\frac{\xi_D-\xi}{4\xi_D-1}
 \Delta  - V(x) \right] \phi^2(x)
 + \left(1 -\frac{D}{2} \right) (P\phi)(x)\phi(x)
 \label{trace''}\:.
\end {eqnarray}
  The same results holds if $\phi(x)\phi(y)$ is replaced by any sufficiently smooth symmetric function 
$\Gamma$.
   Therefore, if $\Gamma$  is as above, similarly to (\ref{trace''}) we get
      \begin{eqnarray}
g^{\mu\nu}(z){\cal D}^{(\eta)}_{(z)\mu\nu} \Gamma(x,y)|_{x=y=z}  &=& \left[\frac{\xi_D-
\xi}{4\xi_D-1}
 \Delta_{(z)}  - V(z) \right] \Gamma(z,z) \nonumber \\
 &+& \left(1 -\frac{D}{2}  + \eta D\right) (P_x\Gamma(x,y))|_{x=y=z}\:.
 \end{eqnarray}
 If $\eta=\eta_D$ and $\Gamma(x,y) = G_\omega^{(1)}(x,y)-Z_{n}(x,y)$, using (\ref{bisolution}) 
and
 Lemma 2.1, we get the   identity in (b).     $-(2c_D/(D+2))U_{D/2}(z,z)$ is the conformal anomaly
 for $V\equiv 0$ and $m=0$, $\xi = \xi_D$ because it coincides with the
 heat-kernel coefficient $a_{D/2}(z,z)/(4\pi)^{D/2}$ \cite{mzp}. This can be seen by direct 
comparison
 of recursive equations defining both classes of coefficients (see references in \cite{mzp}). This 
concludes
 the proof of (b). 
(c) The proof is direct by employing the given definitions.\\
 (d) For $D$ odd the proof of the thesis is trivial. Hence assume $D$ even. In that case, with obvious 
notation,
  $$Z_{\lambda ,n}(x,y)-Z_{\lambda',n}(x,y)= 2\ln\left( \frac{\lambda'}{\lambda}\right) 
\sum_{k=0}^n
c_k s^{k}(x,y)U_{k-1+D/2}(x,y)\:,$$
 where  $c_k$ are numerical coefficients defined above. $\lim_{(x,y)\to (z,z)} {\cal 
D}^{(\eta)}_{(z)(x,y)} (Z_\lambda-Z_{\lambda'})$
 is a polynomial of coefficients $U_{k}(z,z)$ and their derivatives. These coefficients do 
not depend on the
state
  are proportional to heat-kernel ones and thus  are built up as indicated in the thesis  \cite{fulling}. 
Notice that the
obtained tensor field $t$ must be conserved because is the difference of two conserved tensor fields if 
$V'\equiv 0$.
 (e) For $m=0$ the proof of the thesis is trivial because   $Z_{n}$   in Minkoski spacetime does not 
contain the
logarithmic
 term  and does not depend  on both $\lambda$ and $n$, moreover , if $\omega$ is Minkowski
 vacuum $G^{(1)}_\omega(x,y) = Z_{n}(x,y)$  and thus the renormalized stress-energy tensor
 vanishes. Let us consider the case $m>0$.
 In that case, the smooth kernels of the two-point function is  given by, in the sense
 of the analytic continuation if $s(x,y)<0$,
 $$G^{(+)}_\omega(x,y) = \lim_{\epsilon \to 0^+}\frac{4m}{(4\pi)^2
\sqrt{s_{\epsilon,T}(x,y)}}
 K_1\left(m\sqrt{s_{\epsilon,T}(x,y)}\right)$$
with $s_{\epsilon,T}(x,y)\stackrel {\mbox{\scriptsize  def}} {=}s(x,y) + 2 i (T(x)-T(y)) 
+\epsilon^2$ where $\epsilon \to 0^+ $ indicates the
path to approach the branch cut of the squared root along the negative real axis if $s(x,y)<0$.
$T$ indicates any global time coordinate increasing toward the future. $K_1$ is a modified Bessel 
function.
The corresponding Hadamard function can be expandend as
 \begin{eqnarray}
 G^{(1)}_\omega(x,y) &=& \frac{4}{(4\pi)^2 s} + \frac{m^2}{(4\pi)^2}\left\{1 + 
\frac{m^2s}{8}\right\}\ln\left(\frac{e^{2\gamma}m^2s}{4}\right)
+ s^2f(s)\ln\left(\frac{e^{2\gamma}m^2s}{4}\right)\nonumber \\  &-&\frac{m^2}{(4\pi)^2}\left[ 1 
+ \frac{5m^2s}{16} \right] + s^2g(s)  \nonumber
\end{eqnarray}
where $f$ and $g$ are smooth functions and $\gamma$ is Euler-Mascheroni's constant.      Similarly
$$ Z_{\lambda,3}(x,y) =   \frac{4}{(4\pi)^2 s} + \frac{m^2}{(4\pi)^2}\left\{1 + \frac{m^2s}{8} + 
C_2 s^2 + C_3 s^3\right\}\ln\frac{s}{\lambda^2}
\:,$$
where $C_2$ and $C_3$ are constants.   Therefore
\begin{eqnarray}
G^{(1)}_\omega - Z_{\lambda,3}(x,y) &=&
\frac{m^2}{(4\pi)^2}\left\{1 + \frac{m^2s}{8}\right\}\ln\left(\frac{\lambda^2 
e^{2\gamma}m^2}{4}\right)
-   \frac{m^2}{(4\pi)^2}\left[ 1 + \frac{5m^2s}{16} \right]  \nonumber \\
&+& s^2 \left[ h(s) + k(s) \ln s \right] \:,\nonumber
\end{eqnarray}
where $h$ and $k$ are smooth functions.
 Trivial computations lead to
 $$\langle \hat{T}_{\mu\nu}^{(\eta_4)}(z)\rangle_{\omega,\lambda}=  -\frac{m^4}{3(4\pi)^2} \left[
 \ln\left(\frac{\lambda^2 e^{2\gamma}m^2}{4}\right) -\frac{7}{4}\right] g_{\mu\nu}(z)\:.$$
 Posing $\lambda^2 = 4e^{\frac{7}{4}-2\gamma} m^{-2}$ the right-hand side vanishes. $\Box$\\

 \noindent{\em Proof} of {\bf Lemma 3.1.} (a) The thesis can be proved working in 
      a concrete algebra ${\cal W}_\omega(N,{\bf g})$  using operators
      $:\spa K_1\hat\phi \cdots K_n\hat\phi (f)\colon\sp_{H}\in {\cal W}_\omega(N,{\bf g})$
  where $\omega$ is fixed quasifree Hadamard state and $\hat\phi = \hat \phi_\omega$.    
  Notice that, for every $\Psi\in D_\omega$, $\hat{W}_{H, n+1}(x_1,\ldots,x_{n+1})\Psi$ equals
  $$\left(\hat{W}_{H,n}(x_1,\ldots,x_{n}) \hat{W}_{H,1}(x_{n+1})\Psi\right)_S
  - \left(\sum_{l}  \hat{W}_{H,n}(x_1,\ldots,\hat{x}_l,x_{n+1})\Psi H(x_l,x_{n+1})\right)_S$$
  where $_S$ indicates the symmetrization with respect to all arguments and $\hat x_l$ indicates that 
the argument is omitted.
  We prove the thesis,  which is true for $n=1$, by induction.     
 If $H'$ is defined as $H$ but with a different choice of $\psi$ and $\{\alpha_k\}$, for each $\Psi\in 
D_\omega$, 
 the formula above and our inductive hypothesis imply that
   $\left(:\spa K_1\hat\phi \cdots K_{n+1}\hat\phi (f)\colon\sp_{H} -
  :\spa K_1\hat\phi \cdots K_{n+1}\hat\phi (f)\colon\sp_{H'} \right)\Psi$ reduces to
    $$\int_{M^n}
   \left(\sum_{l}  \hat{W}_{H,n}(x_1,\ldots,\hat{x}_l,x_{n+1})\Psi S(x_l,x_{n+1})\right)_S
   {^tK_{n+1}}_{(x_{n+1})} \ldots  \: {^tK_{1}}_{(x_1)}
 f(x_1)   \delta_n(x_1,\ldots x_{n+1}) $$
 where $S(x,y)$ is a smooth function which vanishes for  $x=y$ together  with all of its derivatives.
 As $\hat{W}_{H,n}(x_1,\ldots,\hat{x}_l,x_{n+1})\Psi $ is singular on the diagonal we cannot 
directly conclude 
 that the integral vanishes.
 However, there is sequence of  smooth vector-valued functions $\{ U_j\}$ 
 which tends to the distribution in $(\cdots)_S$
 in the sense of H\"ormander pseudo topology in a closed conic set containing the wave front set of 
the distribution.
   It is simply proven that the smearing procedure of distributions  by means of distributions defined in {\bf 
3.1}
   is continuous in the sense of H\"ormander pseudo topology. Therefore 
 $\left(:\spa K_1\hat\phi \cdots K_{n+1}\hat\phi (f)\colon\sp_{H} -
  :\spa K_1\hat\phi \cdots K_{n+1}\hat\phi (f) \colon\sp_{H'}\right)\Psi$ can be computed as 
  the limit  
   $$\int_{M^n}
   \left(\sum_{l}  U_j(x_1,\ldots,\hat{x}_l,x_{n+1}) S(x_l,x_{n+1})\right)_S
   {^tK_{n+1}}_{(x_{n+1})} \ldots  \: {^tK_{1}}_{(x_1)}
 f(x_1)   \delta_n(x_1,\ldots x_{n+1}) \:,$$
 for $j\to \infty$. However, as each $U_n$ is regular, we can conclude that each term of the sequence 
above vanishes and  
 this proves the thesis because $\Psi\in D_\omega$ is arbitrary. \\(b) By (a) we may assume that the 
distributions 
 $H$ and $H'$ are constructed using the same function $\psi$ and the same sequence of numbers 
$\{\alpha_k\}$.
 Define $L=N\cap N'$ and take $f\in {\cal D}(L)$. Since the convex normal neighborhoods define a 
topological 
 base, $L$ is the union of convex normal neighborhoods. In turn, it implies that the compact set
 $supp f \subset L$ admits a finite covering $\{U_i\}$ made of convex normal neighborhoods 
contained in $L$ and thus
 both in $N$ and $N$. Therefore, if $x,y\in U_i$, the squared 
geodesic distance $s(x,y)$ computed by viewing $x,y$ as elements of $N$
 agrees with the analogue by viewing $x,y$ as elements of $N'$ (also if $N\cap N'$ may not be 
convex normal). 
 By consequence $H$ and $H'$ induce  the same distribution on each $U_i$. $\{U_i\}$ is locally 
finite and thus
 there is a smooth partition of the unity $\{\chi_j\}_j$ subordinate to $\{U_i\}$.
 By linearity we have\\
 $:\spa K_1\phi \cdots K_n\phi (f) \colon\sp_H = \sum_{j}\spa :\spa K_1\phi \cdots K_n\phi (\chi_jf) 
\colon\sp_H
 =\sum_{j}\spa :\spa K_1\phi \cdots K_n\phi (\chi_jf) \colon\sp_{H'} = :\spa K_1\phi \cdots K_n\phi 
(f) \colon\sp_{H'}$
 which concludes the proof.
 $\Box$\\

 \noindent{\em Proof} of part of (b) in {\bf Proposition 3.2.} We want to show that, if $N\subset M$
 is a causal domain  then,  $f_j \to f$ in
  ${\cal D}(N)$ entails $t_n[K_1,\ldots, K_n, f_j] \to t_n[K_1,\ldots, K_n, f]$     in the sense of
   H\"ormander pseudo topology in the conic set $\Gamma_n = \{(x_1,k_1;\ldots ;x_n,k_n)\in 
T^*M^n\setminus \{0\}\:|\: \sum_i k_i =0\}$
   which contains the wave front set of all involved distributions.
  Since there is a coordinate patch covering $N$, $\xi : N\to O$, (normal Riemannian coordinates 
centered on some $p\in N$),
  we can make use of the $\bR^n$-distribution definition  of  convergence (see Definition 8.2.2. in 
\cite{hor1}).
  Therefore, in the following,  $f,f_j$ and $t_n[K_1,\ldots, K_n, f], t_n[K_1,\ldots, K_n^, f_j]$
  have to be  understood as distributions in ${\cal D}'(O)$, and ${\cal D}'(O^n)$ respectively, $O$ 
being an open subset of
  $\bR^4$.     Posing $u_j=  t_n[K_1,\ldots, K_n, f_j]$   and $u= t_n[K_1,\ldots, K_n, f]$,
  we have to show that (1) $u_j \to u$ in ${\cal D}'(O^n)$, and this is trivially true by the given 
definitions since $f_n\to f$ in ${\cal D}(O)$,
   and,   (2),
   \begin{eqnarray}
    \sup_V|k|^N |\widehat{\psi u}(k) -\widehat{\psi u_j}(k)| \to 0, \:\:\:\:\:\mbox{as $j\to +\infty$}        
\label{thesis}
   \end{eqnarray}
   for all $N=0,1,2,\ldots$, $\psi \in {\cal D}(O)$ and $V$, closed conic set\footnote{A conic set 
$V\subset \bR^m$ is
   a set such that if $v\in V$, $\lambda v\in V$ for every $\lambda >0$.} in $\bR^{4n}$ such that
  \begin{eqnarray}\Gamma_n \cap (supp\:\: \psi \times V) = \emptyset \label{defV} \:.\end{eqnarray}
   $\widehat{v}$ denotes the Fourier transform of $v$.
   From now on $K$ denotes a generic vector in $\bR^{4n}$ of the form $(k_1,\ldots,k_n)$, $k_i\in 
\bR^4$ and,
   in components $k_i = (k_i^{1}, k_i^{2}, k_i^{3}, k_i^{4})$.
   We leave to the reader to show that,   with the definitions given above,   $\widehat{\psi u}_{j}$  
and  $\widehat{\psi u}$ are
   polynomials in the components of $K$, whose coefficients smoothly  depend on $k_1+\cdots 
+k_n$, i.e.,
   \begin{eqnarray}\widehat{\psi u}_{j}(k) = \sum_{r_{11},\ldots, r_{n4}\in \bN} 
a_{j r_{11},\ldots, r_{n4}}\left(k_1+\cdots +k_n\right)
   \prod_{i=1}^{n} \prod_{m=1}^4 (k_i^{m})^{r_{im}}\:.\label{start}\end{eqnarray}
   An analogous identity concerning $u$ and coefficients $a_{r_{11},\ldots, r_{n4}}\left(k_1+\cdots +k_n\right)$ 
   holds by omitting $j$  in both sides.
   (Above $0\in \bN$ and only a finite number of functions   $a_{j r_{11},\ldots, r_{n4}}$ and  
   $a_{r_{11},\ldots, r_{n4}}$
   differ from the null function.) 

    Moreover $f_j\to f$ in ${\cal D}(O)$ implies that, for every $N\in \bN$ and $r_{im}\in \bN$,
    \begin{eqnarray}
      \sup_{x\in \bR^4} |x|^N \left | a_{j \:\:\ r_{11},\ldots, r_{n4}} \left( x \right)-
     a_{r_{11},\ldots, r_{n4} }\left(x\right) \right | \to 0      \label{hypothesis2}
    \end{eqnarray}
    With the given notations, our thesis  (\ref{thesis}) reduces to
    \begin{eqnarray}
    \sup_{K\in V} |K|^N \left | \sum_{r_{11},\ldots, r_{n4}\in \bN} \left[a_{j \:\:\ r_{11},\ldots, 
r_{n4}}\left(\sum_{i=1}^4k_i \right)-
     a_{r_{11},\ldots, r_{n4}}\left(\sum_{i=1}^4k_i\right)\right] \prod_{i=1}^{n} \prod_{m=1}^4 
(k_{i}^m)^{r_{im}}
     \right | \to 0      \label{thesis2}
     \end{eqnarray}
    as $j\to +\infty$ for all $N\in \bN$  and $V$ which is closed in $\bR^{4n}$ conic and such that
    \begin{eqnarray}V\cap \{K\in \bR^{4n}\setminus \{0\} \:\:|\:\: \sum_{i=1}^n k_i = 0, \} = 
\emptyset \label{VC}\:.\end{eqnarray}
   We want to prove (\ref{thesis2}) starting from (\ref{hypothesis2}).
   Consider a  linear bijective map $A: K \mapsto Q \in \bR^{4n}$, where $Q=(q_1,\ldots, q_n)$
     and $q_1 = p_1+\cdots + p_n$.  The functions $x \mapsto b_{s_{11},\ldots, s_{n4}}\left(x\right)$ and 
      $x \mapsto b_{j s_{11},\ldots, s_{n4}}\left(x\right)$ which arise
        when translating (\ref{start}) (and the analog for $u$) in the variable $Q$, i.e.,
      $$\widehat{\psi u}_{j}(k) = \sum_{s_{11},\ldots, s_{n4}\in \bN} b_{j s_{11},\ldots, 
s_{n4}}\left(q_1\right)
   \prod_{i=1}^{n} \prod_{m=1}^4 (q_i^{m})^{s_{im}}\:,$$
     are linear combinations of  the functions $x \mapsto a_{j r_{11},\ldots, r_{n4}}\left(x\right)$ 
(with
     coefficients which do not depend on $j$) and {\em vice versa},
     therefore (\ref{hypothesis2}) entails
      \begin{eqnarray}
      \sup_{x\in \bR^4} |x|^N \left | b_{j \:\:\ s_{11},\ldots, s_{n4}} \left( x\right)-
     b_{s_{11},\ldots, s_{n4} }\left(x\right) \right | \to 0      \label{hypothesis3}
    \end{eqnarray}
     for every $N\in \bN$ and $s_{im}\in \bN$.
     Since linear bijective maps transform closed conic sets into closed conic sets and  $|Q| \leq ||A|| 
|K|$, $|K|\leq ||A^{-1}|| |Q|$, our thesis (\ref{thesis2})    is equivalent to
     \begin{eqnarray}
    \sup_{Q\in U} |Q|^N \left | \sum_{s_{11},\ldots, s_{n4}\in \bN}  \left[b_{j \:\:\ s_{11},\ldots, 
s_{n4}}\left(q_1 \right)-
     b_{s_{11},\ldots, s_{n4}}\left(q_1\right)\right] \prod_{i=1}^{n} \prod_{m=1}^4 
(q_{i}^m)^{s_{im}}
     \right | \to 0      \label{thesis3}
     \end{eqnarray}
    as $j\to +\infty$ for all $N\in \bN$  and $U$ closed in $\bR^{4n}$ conic and such that
    \begin{eqnarray}U\cap \{Q\in \bR^{4n}\setminus \{0\} \:\:|\:\:  q_1 = 0\} = \emptyset 
\label{UC}\:.\end{eqnarray}
     It is possible to show that if $U\in \bR^{4n}$ is a set which fulfills (\ref{UC}) and $U$ is {\em 
conic} and  {\em closed} in $\bR^{4n}$, then
    there is $p\in  \bN\setminus \{0\}$
    such that $U\subset U_p$, where       the closed set $U_p$ is defined by
    $$U_p = \left\{Q\in \bR^{4n}\:\:\left |\:\: |q_1| \geq \frac{1}{p} \sqrt{|q_2|^2+\cdots + 
|q_n|^2}\right\}\:,\right. $$
    and thus   $U_p\cap \left\{Q\in \bR^{4n}\setminus \{0\} \:\:|\:\:  q_1 = 0\right\} = \emptyset$   for 
every $p\in \bN\setminus \{0\}$.
    The proof is left to the reader (hint: $U$ is conic  and satisfies (\ref{UC}) then, reducing to a 
compact neighborhood
     of the origin of $\bR^{4n}$, one finds a sequence of points of $U$ which converges to some 
point  $x \in \{Q\in \bR^{4n}\setminus \{0\} \:\:|\:\:  q_1 = 0\}$
     this is not possible because  $\overline{U} = U$ and thus $x\in U\cap \{Q\in \bR^{4n}\setminus 
\{0\} \:\:|\:\:  q_1 = 0\}$).
    If (\ref{thesis3}) holds on each $U_p$, it must hold true on each conic closed set $U$  which 
fulfills (\ref{UC}).
    The validity of  (\ref{thesis3}) on each $U_p$ is a direct consequence of (\ref{hypothesis3}) and 
the inequalities which holds on
    $U_p$
    $$ |Q| \leq \sqrt{1+p^2} |q_1| \:\:\: \: \mbox{and}\:\:\:\:  |q_r^s| \leq |q_1|/p\:\:\:\:\:\mbox{for 
$r=2,3,\ldots,n$, $s=1,2,3,4$. 
    $\Box$}$$

 \end{document}